\documentclass[a4paper,fleqn,usenatbib]{mnras}
\usepackage{dcolumn}
\usepackage{graphicx}
\usepackage{url}
\usepackage{color}
\usepackage[T1]{fontenc}
\usepackage{ae,aecompl}
\usepackage{pdflscape}
\usepackage{epsfig,color}

\newcommand{\cm}{cm$^{-1}$}

\newcommand{\Ai}{\textit{Ab initio}}
\newcommand{\etal}{\textit{et al.}}

\newcommand{\eqref}[1]{(\ref{#1})}

\title[ExoMol XVI: Line list for H$_2$S]{ExoMol molecular line lists  - XVI: The rotation-vibration spectrum of hot H$_2$S}

\author[A.A.A. Azzam et al.]{Ala'a A. A. Azzam$^{1,2}$,  Jonathan Tennyson$^{1}$\thanks{Email: j.tennyson@ucl.ac.uk}, Sergei N. Yurchenko$^{1}$,
\newauthor Olga V. Naumenko$^{3}$
\\\
$^{1}$Department of Physics and Astronomy, University College London, London WC1E 6BT, UK\\
$^{2}$Department of Physics, The University of Jordan, Queen Rania Street, Amman 11942, Jordan\\
$^{3}$Institute of Atmospheric Optics, Russian Academy of Sciences, Tomsk, Russia}
\date{Accepted XXXX. Received XXXX; in original form XXXX}
\pubyear{2016}

\begin{document}
\label{firstpage}
\pagerange{\pageref{firstpage}--\pageref{lastpage}}

\maketitle

\begin{abstract}

    This work presents the AYT2 line list: a comprehensive list of 114 million
 $^{1}$H$_2$$^{32}$S vibration-rotation transitions computed using
  an empirically-adjusted potential energy surface and an {\it ab
    initio} dipole moment surface. The  line list  gives complete
  coverage up to 11000 \cm\ (wavelengths longer than 0.91 $\mu$m) for
  temperatures up to 2000 K. Room temperature spectra can be
simulated up to 20000 \cm\ (0.5 $\mu$m) but the predictions at
visible wavelengths are less reliable. AYT2 is made available in
  electronic form as supplementary data to this article and at
  \url{www.exomol.com}.
\end{abstract}


\section{Introduction}
The investigation of the sulphur chemistry in space is a subject of the active research
\citep{01RuKixx.SO,04WaCaCe.H2S,06ViLoFe.H2S,09ZaMaFr.H2S,11AlMaMa.H2S,13ReSeBa.H2S}. In particular
\citet{13ReSeBa.H2S} studied the atmospheric
composition and the spectra of earth-like exoplanets with sulphur
compounds such as hydrogen sulphide (H$_{2}$S) and sulphur dioxide (SO$_{2}$)
using a one-dimensional photochemistry model and associated radiative
transfer model to investigate sulphur chemistry in atmospheres
ranging from reducing to oxidising.
\citet{06ViLoFe.H2S} used thermochemical equilibrium and kinetic calculations to model sulphur chemistry in giant
planets, brown dwarfs, and extrasolar giant planets, and found that H$_2$S
is the dominant S-bearing gas throughout substellar atmospheres and approximately represents the atmospheric
sulphur inventory. Therefore,
observations of H$_{2}$S in these objects should provide a good estimate of their atmospheric sulphur content.
H$_2$S has been, however, ruled out as a potential biosignature in atmospheres of exoplanets
according to a biomass-based model study by \citet{13SeBaHu.exoplanet}.

H$_2$S has long been known in the interstellar medium
\citep{72ThWiKu.H2S} and is important in star-forming
\citep{04WaCaCe.H2S,15NeGoGe.H2S} and circumstellar \citep{
  93OmLuMo.H2S} regions.  \citet{11AlMaMa.H2S} detected H$_{2}$S for
the first time in galaxy M82, where they studied the chemical
complexity towards the central parts of the starburst galaxy, and
investigated the role of certain molecules as tracers of the physical
processes in the galaxy circumnuclear region.  \citet{01RuKixx.SO}
found evidence for SO$_{2}$, SO and H$_{2}$S sulphide in Io's
exosphere.  For Venus, the H$_{2}$S composition of the atmosphere at
altitudes below 100 km was studied by \citet{85ZaMoxx.H2S} and
\citet{06BeMoTa.H2S}.  Determination of the abundances of gases such
as CO, SO$_{2}$, OCS, S$_{2}$ or H$_{2}$S near the surface is
important to constrain the oxidation state of the lower atmosphere and
surface, and determine the stability of various minerals. Also,
measurements at higher altitudes of, for example, SO$_{3}$, SO or
elemental sulphur, are needed to better understand the sulphur cycle
and the chemistry at work below the cloud base. Conversely a recent
search for H$_2$S in volcanic emissions on Mars failed to detect any
\citep{15KhViMu}.  H$_2$S is known to be present in comets
\citep{02BiBoCr.comets} being first detected by
\citet{91BoCoCr.comets}.

On Earth naturally occurring H$_2$S is associated with volcanic
activity \citep{12HoHoLa.H2S}.  Gaseous H$_2$S is also detected in a
number of other situations including emissions from waste water
\citep{12LlEsMa.H2S} and as a by-product of industrial processes
\citep{13SzMoGu.H2S}.

Known experimental absorption spectra of H$_2$S molecule cover the
region from the microwave up to the visible (0.6 $\mu$m).
Observations include transitions belonging to 59 vibrational bands
associated with different 14 polyads \citep{12PoLaVo.H2S}, where the
polyad number is defined as $n=v_{1}+v_{2}/2+v_{3}$, where $v_{i}$ are
standard normal-mode vibrational quantum numbers.

\begin{figure}
\centering
{\leavevmode \epsfxsize=9.0cm \epsfbox{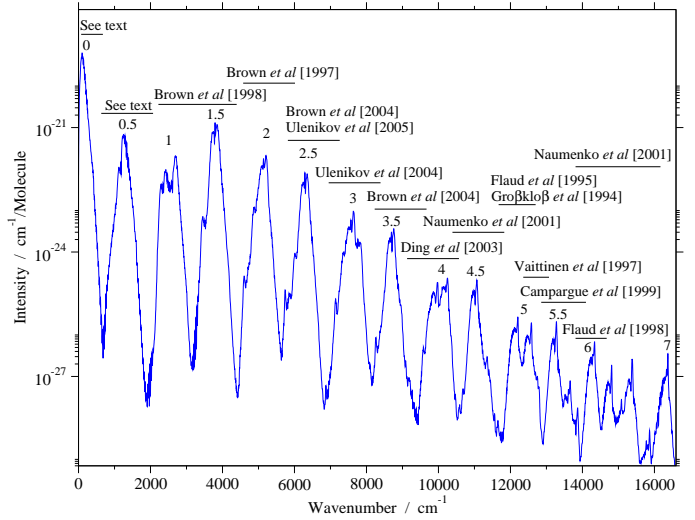}}
\caption{Summary of experimental work on the H$_{2}$S absorption spectrum overlayed with our theoretical cross-sections. All spectra were recorded at room temperature. The polyad number of each band is also given.}
\label{summary-for-experimenatl-work-on-H2S}
\end{figure}

The rotational band has received attention from many experimentalists
\citep{53BuJRGo.H2S,66HuDyxx.H2S,68CuKeGa.H2S,69MiLeHa.H2S,71Huxxxx.H2S,72HeCoDe.H2S,83FlCaJo.H2S,85BuFeMe.H2S,94YaKlxx.H2S,95BeYaWi.H2S,jt558,14CaPuGa.H2S}.
The first bending vibrational band ($\nu_2$) at 1183~cm$^{-1}$ was
studied by \citet{82LaEdGi.H2S}, \citet{83Stxxxxa.H2S} and
\citet{96UlMaKo.H2S}.  The two fundamental stretching; symmetric
($\nu_1$) and asymmetric ($\nu_3$) lying at 2615 and 2626 cm$^{-1}$,
respectively, are not isolated but overlapped with strong Coriolis and
Fermi resonance interactions.  The first triad region ($2\nu_2,\nu_1,$
and $\nu_3$) was studied by \citet{81GiEdxx.H2S}, the second triad
region ($3\nu_2,\nu_1+\nu_2,$ and $\nu_2+\nu_3$) was studied by
\citet{69SnEdxx.H2S} and \citet{96UlOnKo.H2S}, while
\citet{98BrCrCr.H2S} studied these two triad regions simultaneously.
The 4500~--~5600 cm$^{-1}$ spectral region was investigated by
\citet{97BrCrCr.H2S}.  \citet{04BrNaPoa.H2S} and \citet{05UlLiBe.H2S}
recorded and analysed the transitions in the region 5700~--~6600~\cm.
In the 7300~--~7900~\cm\ region, more than 1550 transitions up to
$J=14$ were recorded and analysed by \citet{04UlLiBe.H2S}.  The
absorption spectrum in the region 8400~--~8900~\cm\ was recorded by
\citet{04BrNaPob.H2S}. \citet{94ByNaSm.H2S} recorded and analysed
spectra between 2000 to 11 147 \cm.  A number of shorter wavelength
regions have been studied, namely, 9540~--~10~000~\cm\ by
\citet{03DiNaHu.H2S}, 10~780~--~11~330~\cm\ by \citet{01NaCaxxb.H2S},
11~930~--~12~300~\cm\ by \citet{94GrRaSt.H2S} and
\citet{95FlGrRa.H2S}, 12~270~--~12~670~\cm\ by \citet{97VaBiCa.H2S},
near 13~200 \cm\ by \citet{99CaFlxx.H2S}, 14~100~--~14~400~\cm\ by
\citet{98FlVaCa.H2S}, and 16~180~--~16~440~\cm\ by
\citet{01NaCaxxa.H2S}. This situation is summarised in
Fig.~\ref{summary-for-experimenatl-work-on-H2S}.

All this work has been performed using cool samples, that is below 300
K.  The highest recorded value of the rotational quantum number $J$ is
22 in the rotational band region, and the highest predicted value is
27 in the same region. Altogether around 10~000 ro-vibrational energy
levels are known from these experiments.  The spectroscopic data for
the H$_{2}$S molecule has been used to populate various spectroscopic
databases.  Table~\ref{databases} summarises the contents of the
HITRAN-2012 \citep{jt557}, GEISA \citep{jt504,jt636}, W@DIS
\citep{12PoLaVo.H2S}, CDMS \citep{01MuThRo.db,05MuScSt.db}, and JPL
\citep{98PiPoCo} databases.  All these databases contain data
resulting from fitted effective Hamiltonians, apart from W@DIS which
contains only measured transitions without intensities.  The 2012
release of HITRAN \citep{jt557} updated HITRAN 2008 \citep{jt453}
using extra rotational data rotational band of H$_{2}$S spectrum from
\citet{jt558} and the data published in IAO LMS Spectra
[spectra.iao.ru].

While absorption spectra of H$_2$S at elevated temperature have recently been
recorded in the ultra-violet by \citet{15GrFaCl.H2S},
we are unaware of any high resolution experimental studies of H$_2$S infrared spectra at higher
than ambient temperature.
H$_2$S cross-sections have been measured up to  $T = 50$~C  as part of PNNL  database \citep{PNNL}.

\begin{table*}
\begin{center}
\caption{Summary for the H$_{2}$S spectral data available in different databases.}
\label{databases}
\begin{tabular}{c c r r r r}
\hline
\hline
    Database   &    isotopologue &\multicolumn{1}{c}{Bands}   &\multicolumn{1}{c}{Transitions}   &\multicolumn{1}{c}{Wavenumber$_{min}$}& \multicolumn{1}{r}{Wavenumber$_{\rm max}$}\\
               &                 & \multicolumn{1}{c}{\#}     & \multicolumn{1}{c}{\#}           &                  (\cm)               &                    (\cm)              \\
\hline
HITRAN 2012    & H$_{2}$$^{32}$S &            49             &               36561            &             2                     & 11330  \\
               & H$_{2}$$^{33}$S &            19             &                6322           &             5                     & 11072 \\
               & H$_{2}$$^{34}$S &            24             &               11352             &             5                     & 11227  \\
GEISA          & H$_{2}$$^{32}$S &            14             &               12330            &             2                     & 4257  \\
               & H$_{2}$$^{33}$S &             8             &                3564            &             5                     & 4098  \\
               & H$_{2}$$^{34}$S &             8             &                4894            &             5                     & 4171  \\
  W@DIS        & H$_{2}$$^{32}$S &            59             &               34148            &             1                    & 16437   \\
  CDMS         & H$_{2}$$^{32}$S &             1             &                1501            &             1                     & 554   \\
               & H$_{2}$$^{33}$S &             1             &                4759            &             1                     & 402    \\
               & H$_{2}$$^{34}$S &             1             &                 990            &             1                     & 444    \\
  JPL          & H$_{2}$$^{32}$S &             1             &                1525            &             1                     & 333    \\
\hline
\hline
\end{tabular}
\end{center}
\end{table*}

A number of theoretical studies have considered H$_{2}$S. Its ro-vibrational spectrum  was calculated
by \citet{89SeCaZi.H2S}, \citet{jt266} and \citet{04TyReSc.H2S}.
In the work of \citet{89SeCaZi.H2S}, the room temperature absorption ro-vibrational spectrum of H$_{2}$S was
calculated variationally for the pure rotational
and $\nu_{2}$, $2\nu_{2}$, $\nu_{1}$, and $\nu_{3}$ transitions from $J = 0$ to 13, where the full account of
the anharmonicity effects and ro-vibration couplings were considered. \citet{89SeCaZi.H2S} calculated the vibrational band
origins of the fundamental transitions with accuracy better
than 10~\cm, and the ro-vibrational transitions to within a few tenths of a \cm~for low $J$'s and up to a few \cm~for
high $J$'s; their work was also extended to deuterated isotopologues \citep{jt99}.
The anomalies in the spectral intensity
for this molecule were obtained qualitatively.
\citet{jt266} calculated the vibrational band origins of H$_{2}$$^{32}$S with accuracy of 29~\cm~up to 14~300~\cm, and the
rotational transitions of the ground
vibrational state for $J = 17$ with deviations from the experimental values from 2 to 10~\cm.
\citet{04TyReSc.H2S} used a spectroscopically-determined potential energy surface to compute the spectrum in the interval 0~--~8000~\cm,
for $J$ up to 18 and  with and
intensity cut-off $\leq~10^{-27}$~cm$^{-1}$/(molecule$\times$cm$^{-2}$); they reported
calculated transitions to be better
than 0.01~\cm~for the line positions and 1-3\% in the intensities for the strong and medium
lines and to $\sim$10\% for the weak lines for room temperature conditions. Very recently \citet{15CaLexx.H2S}
explored the  use of ladder operators to study the vibrational spectrum of this system.

Remote detection of H$_{2}$S relies on well-characterised laboratory spectra.
At higher temperatures the resulting list of transitions becomes very extensive
and is best calculated using a robust theoretical model \citep{jt511}.
The ExoMol project \citep{jt528} aims to provide molecular line lists
for exoplanet and other atmospheres with a particular emphasis on
hot species.  In this work we present a comprehensive, hot linelist of  vibration-rotation transitions
of $^{1}$H$_2$$^{32}$S. This line list
should be appropriate  for temperatures up to
2000~K. The methodology used, which is discussed in the following section, closely
follows that used to generate comprehensive hot line lists for triatomic
species such as water \citep{jt378,jtpoz} and, recently, SO$_2$ \citep{jt635}.
Section 3 presents our line list computations. Results
and comparisons are given in section 4. Section 5 gives our
conclusions.

\section{Theoretical method}

In order to compute a line list for H$_2$S three things are required \citep{jt475}:
a suitable potential energy surface (PES), dipole moment surfaces
(DMS), and a nuclear motion program. Each of these are considered
in turn below.

\subsection{Potential Energy Surfaces}
Four PESs were tested. Our starting point was an {\it ab initio} PES constructed using the CCSD(T)/aug-cc-pV(Q+d)Z
level of theory as implemented in {\sc Molpro} \citep{12WeKnKn.methods}. This surface was fitted
to the function form given by \citet{01TyTaSc.H2S} using 1200 geometries
covering the energy range up to 40~000 cm$^{-1}$ above equilibrium. The surface was then refined
by fitting it to the available experimental values
of H$_2$S for $J$~$\leq$~6 covering the energy range up to 16~500 cm$^{-1}$ with a root-mean-square (rms) error of 0.03 cm$^{-1}$ for
the fit. We
will call this surface PES-Y. This surface was tested by calculating the energy levels for $J$~=~0, 1, 2, 5 and 10, and
comparing the results with experimental energy levels.

The second PES, PES-T, was  constructed by \citet{01TyTaSc.H2S},
using a dataset of then-available experimentally determined H$_2$S energy levels.
This surface was obtained by the simultaneous fit of a large sample of high-resolution ro-vibrational data, using
an extensive set of more than 12~000 experimental ro-vibrational transitions for 7 isotopologues of H$_2$S.
This surface is the most accurate available empirically-determined, PES. However, using the published parameters of PES-T
and calculating the ro-vibrational
energy levels, we found some problems.
Tyuterev {\etal} used {\sc Dvr3d} \citep{jt338} to confirm the convergence of the basis set used in their work for high vibrational states.
First, calculating the
vibrational energy levels using PES-T with
the parameters suggested for {\sc Dvr3d} by Tyuterev {\etal}
($r_e$~=~2.75, $D_e$~=~0.1, $\omega_e$~=~0.01, all in a.u., NPNT2~=~35 and NALF~=~98), and then
comparing these calculated energy values with the experimental vibrational energy levels taken from \citet{01TyTaSc.H2S},
we could not reproduce the values for the vibrational energy
levels as published by Tyuterev {\etal}.
Second, testing PES-T for convergence shows that increasing the number of the radial points, NPTN2, some of the
energy levels become negative. Graphical investigations showed that this surface develops a hole when the atoms all lie close together.
This problem was solved by considering the coefficients up to the quadratic order and ignoring
the coefficients with the higher orders in the refining function for the energies above 50~000~cm$^{-1}$. The modified PES-T was used
to calculate the vibrational energy levels again. A further problem is  that we found
Tyuterev~{\etal}'s vibrational basis set is not converged above 9000~\cm.

PES-T was further refined using updated experimental levels lying
up to 17~000 cm$^{-1}$; 71 parameters were fitted in two different refinements:
(1) using experimental energy levels with $J$ = 0, 1, 2 and 5; (2) using experimental energy levels with
$J$ $\leq$ 6. The resulting two PESs will be referred to as PES-Y0125 and PES-Y0-6, respectively.
Convergence tests were performed also for these new refined surfaces by calculating
ro-vibrational energy levels for $J$ = 0, 1, 2, 5 and 10, and comparing the calculated values with the available experimental data.
PES-Y0125 gives better results than PES-Y0-6.

PES-Y0125 predicts experimentally known energy levels with $J$ $\le$
10 with a standard deviation of 0.11 cm$^{-1}$ compared to 0.23
cm$^{-1}$ using PES-T (using our parameters for {\sc Dvr3d}). Note that Tyuterev
{\etal} claimed that their PES predicted the experimentally known levels
with $J$ $\le$ 15 with a standard deviation of 0.03 cm$^{-1}$
for all isotopologues.
Table~\ref{sd} shows the standard deviations for the calculated
ro-vibrational energy levels up to 17~000 cm$^{-1}$ using PES-T and
PES-Y0125 for $J$ = 0, 1, 2, 5 and 10.  These calculations show that
using PES-Y0125, around 7\%\ of the ro-vibrational energy level with $J \leq 5$
values have errors more than 0.25 cm$^{-1}$. All of these levels lie above
12~450 cm$^{-1}$. This proportion increases
with $J$ so that 26\%\ of levels are  more than 0.25 cm$^{-1}$ away
from the observed
$J$~=~10 levels, all of them above 8600 cm$^{-1}$. PES-Y0125 was
adopted for this study.

\begin{table}
\begin{center}
\caption[]{Comparison between the experimental and the calculated vibrational levels in \cm.
Columns 1 and 2 give quantum numbers in normal and local modes.
Column 3 gives observed vibrational bands origins, see  \citet{01TyTaSc.H2S};
column 4 gives the residuals of these bands origins published by \citet{01TyTaSc.H2S};
column 5 gives the residuals of the bands origins computed using PES-T with
the {\sc Dvr3d} parameters  suggested by \citet{01TyTaSc.H2S};
column 6 gives residuals of the bands origins computed using PES-Y0125 and
our parameters for {\sc Dvr3d}.}
\label{compa}
\scalebox{0.8}{
\begin{tabular}{c c r r r r}
\hline\hline
 Normal                & Local              &\multicolumn{1}{c}{Obs.}     &\multicolumn{3}{c}{Obs.-Calc.}                 \\ \cline{4-6}
 ($\nu_1$$\nu_2$$\nu_3$)&[$n_1$ $n_3^\pm$,b]& &4&5&6\\

\hline
                       &                    &                             &                                       &                                     &                                     \\
(010)                  &[00$^+$,1]          &1182.58                      &-0.01                                  & -0.01                               & 0.02                                \\
(020)                  &[00$^+$,2]          &2353.96                      & 0.00                                  & -0.01                               &-0.02                                \\
(100)                  &[10$^+$,0]          &2614.41                      &-0.02                                  & -0.01                               & 0.06                                \\
(030)                  &[00$^+$,3]          &3513.79                      & 0.00                                  &  0.00                               &-0.13                                \\
(110)                  &[10$^+$,1]          &3779.17                      & 0.00                                  &  0.00                               &-0.11                                \\
(040)                  &[00$^+$,4]          &4661.68                      &-0.01                                  &  0.00                               &-0.35                                \\
(120)                  &[10$^+$,2]          &4932.70                      & 0.01                                  &  0.01                               &-0.16                                \\
(200)                  &[20$^+$,0]          &5144.99                      &-0.01                                  &  0.02                               & 0.07                                \\
(002)                  &[11$^+$,0]          &5243.10                      &-0.02                                  & -0.01                               & 0.05                                \\
(050)                  &[00$^+$,5]          &5797.24                      & 0.01                                  &  0.01                               &-0.69                                \\
(130)                  &[10$^+$,3]          &6074.58                      &-0.04                                  & -0.04                               &-0.23                                \\
(210)                  &[20$^+$,1]          &6288.15                      & 0.04                                  &  0.07                               & 0.05                                \\
(102)                  &[30$^+$,0]          &7576.38                      &-0.03                                  &  0.04                               & 0.02                                \\
(300)                  &[21$^+$,0]          &7752.26                      &-0.01                                  &  0.02                               & 0.15                                \\
(112)                  &[30$^+$,1]          &8697.14                      &-0.01                                  &  0.06                               & 0.08                                \\
(202)                  &[40$^+$,0]          &9911.02                      & 0.00                                  &  0.18                               & 0.04                                \\
(400)                  &[31$^+$,0]          &10188.30                     &-0.01                                  &  0.06                               & 0.12                                \\
(212)                  &[40$^+$,1]          &11008.68                     &-0.05                                  &  0.13                               &-0.02                                \\
(302)                  &[50$^+$,0]          &12149.46                     & 0.04                                  &  0.40                               & 0.21                                \\
(104)                  &[41$^+$,0]          &12524.63                     & 0.01                                  &  0.19                               & 0.07                                \\
(312)                  &[50$^+$,1]          &13222.77                     & 0.05                                  &  0.42                               &-0.22                                \\
(322)                  &[50$^+$,2]          &14284.71                     & 0.02                                  &  0.42                               &-0.74                                \\
(402)                  &[60$^+$,0]          &14291.12                     & 0.03                                  &  0.66                               & 0.58                                \\
\hline\hline
\end{tabular}
}
\end{center}
\end{table}

\begin{table}
\begin{center}
\caption{Standard deviation values of the ro-vibrational energy levels up to 17~000 cm$^{-1}$ for
$J$ = 0, 1, 2, 5 and 10.}
\label{sd}

\begin{tabular}{c c c}
\hline\hline
  &\multicolumn{2}{c}{\textbf{\textbf{Standard deviation (cm$^{-1}$)}}} \\
  \cline{2-3}
$J$&          PES-Y0125   &     PES-T                                   \\
\hline
   &                      &                                              \\
0  &           0.19       &      0.24                                    \\
1  &           0.06       &      0.21                                    \\
2  &           0.07       &      0.21                                    \\
5  &           0.07       &      0.23                                    \\
10 &           0.19       &      0.24                                    \\
\hline\hline
\end{tabular}
\end{center}
\end{table}

\subsection{Dipole Moment Surfaces}

The preferred method of generating accurate intensities is to use {\it
  ab initio} DMS \citep{jt156,jt573,jt613}.  However H$_2$S
intensities are known to be particularly difficult to reproduce since
they display a number of anomalies. For example, the observed
intensity of all the fundamental bands of H$_2$S are two-to-three
orders of magnitude weaker than those in similar triatomics such as
H$_2$O and H$_2$Se, and are also much weaker than those of combination
bands ($\nu_{1}+\nu_{2}$, $\nu_{2}+\nu_{3}$ and $\nu_{1}+\nu_{3}$)
\citep{98BrCrCr.H2S}. In particular the usually strong asymmetric
stretch fundamental band $\nu_{3}$ is even weaker than the $2 \nu_{2}$
bending overtone. Furthermore all fundamental bands show intensity
anomalies in their rotational distributions
\citep{98BrCrCr.H2S,81GiEdxx.H2S,83Stxxxxa.H2S}. For the $\nu_{3}$
band, some `forbidden' $\Delta K_{a} = \pm 2$ transitions are actually
more intense than the corresponding `allowed' $\Delta K_{a} = 0$
transitions \citep{98BrCrCr.H2S}.  Reproducing this behaviour {\it ab
  initio} therefore represents a considerable challenge. \Ai\ DMS have
been calculated for H$_{2}$S by \citet{89SeCaZi.H2S}, by Cours,
Tyuterev and co-workers \citep{02CoRoTy.H2S,00CoRoTy.H2S,03HeCoTy.H2S}
and us \citep{jt607}.

In this work we use the ALYT2 DMS of \citet{jt607} which uses
a CCSD(T)/aug-cc-pV(6+d)Z level of theory
supplemented by a core-correlation/relativistic corrective surface obtained
at the CCSD[T]/aug-cc-pCV5Z-DK level.
The intensities  computed with this surface agree
to within 10 \%\ when compared with directly measured
experimental data. Further details can be found in \citet{jt607}.

\subsection{Nuclear motion calculations}

Ro-vibrational spectra were computed using the {\sc Dvr3dr} program
suite \citep{jt338} in Radau coordinates and a bisector embedding.
Analysis showed that the published version {\sc Dvr3dr} used a very
significant amount of time constructing the final Hamiltonian matrix
in module {\sc rotlev3b}.  {\sc rotlev3b} uses vibrational functions
generated in the first step of the calculation \citep{jt46} to provide
basis functions for the full ro-vibrational calculation performed by
{\sc rotlev3b}. For high $J$ calculations this algorithm involves
transforming large numbers of off-diagonal matrix elements to the
vibrational basis set representation, see Eq.~(31) in \citet{jt114}.
This step can be refactored as
two successive summations rather than a double summation.  The savings
in doing this proved to be very significant, so much so that computing
the hot ATY2 linelist was actually much quicker than generating the
original (unpublished) small ATY1 room temperature line list.

Considerable care was taken to ensure convergence of the final
calculations, see \citet{13Azzamx.H2S} for details.
Table~\ref{tab:dvr3d} gives the parameters used in the final
calculation which is sufficient to converge all energy levels consider
to about 0.2 \cm, and very much better than this for the vast majority
of them.

\begin{table}
\caption{Input parameters for DVR3DRJZ and ROTLEV3B modules of DVR3D \citep{jt338}.}\label{tab:dvr3d}
\tiny
\begin{tabular}{lll}
\hline\hline
Parameter & Value & Description\\
\hline
 \multicolumn{3}{l}{DVR3DRJZ}\\
NPNT2 & 40 & No. of radial  DVR points (Gauss-Laguerre)\\
NALF & 48 & No. of angular DVR points (Gauss-Legendre)\\
NEVAL &2000 & No. of eigenvalues/eigenvectors required\\
MAX3D & 6000 & Dimension of final vibrational Hamiltonian\\
XMASS (S) & 31.972071 Da & Mass of sulphur atom\\
XMASS (H) & 1.007825 Da& Mass of oxygen atom\\
$r_e$ & 3.8~a$_0$ & Morse parameter (radial basis function)\\
$D_e$ & 0.4~E$_h$ & Morse parameter (radial basis function)\\
$\omega_e$ & 0.005~a.u. & Morse parameter (radial basis function)\\
\hline
 \multicolumn{3}{l}{ROTLEV3B}\\
NVIB & 1400 & No. of vib. functions used for  each $K$\\
\hline
\hline
\end{tabular}
\end{table}

\section{Line list calculations}
\label{linelist calculation procedures}

The calculations for the ATY2 were performed using 16 processors on the machine Amun which are
Intel(R) Xeon(R) CPU E7340 @ 2.40GHz. All states with $J \leq 40$
lying up to 21~000 \cm\ above the vibrational ground state were included.
Einstein A-coefficients were generated by considering all transitions involving a lower state
below 10~000 \cm. As discussed below this means that the AYT2 line list is complete at higher
temperatures for transition wavenumbers up to 11~000 \cm. At lower temperatures it is complete to higher wavenumbers, about
20~000 \cm\ at 296~K. The full line list has
 113~945~193 lines sorted by frequency and split into 20 files in 1000 \cm\ portions.

 Table~\ref{tab:states} gives a portion of the H$_2$S states file.
 {\sc DVR3D} does not provide approximate quantum numbers: $K_a$,
 $K_c$ or the normal mode vibrational labels $\nu_1$, $\nu_2$ and
 $\nu_3$.  Lower-lying levels were checked and assigned quantum
 numbers on the basis of comparison with effective Hamiltonians.
 For higher levels below about 16~000 \cm, labels
 were taken from calculations performed with {\sc
 TROVE} \citep{07YuThJe.method} and used the correlation between {\sc DVR3D} and
 {\sc TROVE}.  The higher stretching
 states of H$_2$S are actually better represented by local mode
 \citep{12Jexxxx.cluster} rather than normal mode quantum numbers.
 However, since there is a one-to-one correspondence between these two
 representations for XH$_2$ molecules \citep{jt242}, we simply use the
 normal mode representation.  We note that these quantum numbers are
 approximate and may be updated in future as better estimates become
 available.  The table also includes a column which gives the
 calculated radiative decay lifetime for each state. This is a new
 feature in the ExoMol file structures \citep{jt631}.
 Table~\ref{tab:trans} gives a portion of the ATY2 transitions file.

\begin{table}
\caption{ Extract from the state file for H$_2$S. The full table is available from
http://cdsarc.u-strasbg.fr/cgi-bin/VizieR?-source=J/MNRAS/xxx/yy.}
\tiny
\renewcommand{\arraystretch}{0.9}
\begin{tabular}{rrrrrrrrrrr}
\hline\hline
$i$ & $\tilde{E}$& $g$ & $J$ & $\tau$ & $\Gamma$ & $K_a$ & $K_c$ & $\nu_1$ & $\nu_2$ & $\nu_3$ \\
\hline
  1  &        0.000000  & 1  & 0  & \verb!    inf    !&    A1   &    0  & 0  & 0  & 0  &     0  \\
  2  &     1182.569618  & 1  & 0  & \verb! 2.2209E+01!&    A1   &    0  & 0  & 0  & 1  &     0  \\
  3  &     2353.907317  & 1  & 0  & \verb! 9.9513E+00!&    A1   &    0  & 0  & 0  & 2  &     0  \\
  4  &     2614.394829  & 1  & 0  & \verb! 1.5263E+01!&    A1   &    0  & 0  & 1  & 0  &     0  \\
  5  &     3513.705072  & 1  & 0  & \verb! 5.1249E+00!&    A1   &    0  & 0  & 0  & 3  &     0  \\
  6  &     3779.189348  & 1  & 0  & \verb! 1.0827E+00!&    A1   &    0  & 0  & 1  & 1  &     0  \\
  7  &     4661.605794  & 1  & 0  & \verb! 2.7037E+00!&    A1   &    0  & 0  & 0  & 4  &     0  \\
  8  &     4932.688937  & 1  & 0  & \verb! 5.7855E-01!&    A1   &    0  & 0  & 1  & 2  &     0  \\
  9  &     5145.031868  & 1  & 0  & \verb! 2.2070E+00!&    A1   &    0  & 0  & 2  & 0  &     0  \\
 10  &     5243.158956  & 1  & 0  & \verb! 2.7018E+00!&    A1   &    0  & 0  & 0  & 0  &     2  \\
 11  &     5797.207552  & 1  & 0  & \verb! 1.4956E+00!&    A1   &    0  & 0  & 0  & 5  &     0  \\
 12  &     6074.566059  & 1  & 0  & \verb! 3.9968E-01!&    A1   &    0  & 0  & 1  & 3  &     0  \\
 13  &     6288.134723  & 1  & 0  & \verb! 4.0625E-01!&    A1   &    0  & 0  & 2  & 1  &     0  \\
 14  &     6385.320930  & 1  & 0  & \verb! 3.4910E-01!&    A1   &    0  & 0  & 0  & 1  &     2  \\
 15  &     6920.081316  & 1  & 0  & \verb! 8.8597E-01!&    A1   &    0  & 0  & 0  & 6  &     0  \\
 16  &     7204.435162  & 1  & 0  & \verb! 3.0530E-01!&    A1   &    0  & 0  & 1  & 4  &     0  \\
\hline\hline
\end{tabular}
\label{tab:states}
\mbox{}
{\flushleft
$i$:   State counting number.     \\
$\tilde{E}$: State energy in \cm. \\
$g$: State degeneracy.\\
$J$: Total angular momentum            \\
$\tau$: State lifetime in s$^{-1}$, see  \citet{jt624}.\\
$\Gamma$: Symmetry.\\
$\nu_1$:   Symmetric stretch quantum number. \\
$\nu_2$:   Bending quantum number. \\
$\nu_3$:   Asymmetric stretch quantum number. \\
$K_a$: Asymmetric top quantum number.\\
$K_c$:  Asymmetric top quantum number.\\
}

\end{table}

\begin{table}
\caption{Extract from the transitions file for H$_2$S.
The full table is available from
http://cdsarc.u-strasbg.fr/cgi-bin/VizieR?-source=J/MNRAS/xxx/yy.}
\begin{tabular}{rrr}
\hline\hline
$f$ & $i$ & $A_{fi}$ \\
\hline
       54311   &     54310& 8.2902e-04\\
       96461   &     90862& 7.0642e-10\\
       95160   &     95159& 2.9992e-04\\
      182973   &    182972& 6.2635e-13\\
       66321   &     66320& 8.0872e-08\\
       63788   &     54387& 1.4876e-09\\
      166858   &    166857& 4.1485e-09\\
       66292   &     66291& 1.4730e-07\\
      118595   &    119927& 3.3674e-08\\
       12533   &      9738& 2.2129e-02\\
       67918   &     67917& 2.2181e-02\\
       19633   &     14097& 7.6778e-05\\
       44469   &     44468& 1.8778e-12\\
       41993   &     41992& 2.5145e-06\\
       64167   &     58776& 8.1489e-13\\
       49157   &     41344& 3.7880e-01\\
       44472   &     39869& 3.7880e-01\\
      183309   &    183308& 1.2889e-11\\
       49527   &     49526& 3.3570e-10\\
       86003   &     80100& 9.6741e-12\\
       31768   &     31767& 1.8568e-04\\
\hline\hline
\end{tabular}
\label{tab:trans}

\noindent
 $f$: Upper  state counting number;\\
$i$:  Lower  state counting number;\\
$A_{fi}$:  Einstein-A coefficient in s$^{-1}$.

\end{table}

\subsection{Room temperature comparisons}
\label{Room temperature linelist accuracy}

The accuracy of the calculated spectrum can be checked by comparing the transition positions and
intensities with their counterparts in the literature and the databases. The only available experimental data for
H$_{2}$S spectrum is at room temperature ($T = 296$ K), so we judge the accuracy of our calculations using the spectrum calculated
at this temperature. This is not always straightforward
for two reasons. First, in order to compare spectra in detail,  full assignments are
required for the transitions  and this is not the case for
the calculated transitions using {\sc Dvr3d}.
Second, our dipole study \citep{jt607}
showed that agreement between our calculations and experimentally measured transitions is much better
than  with transitions which are the result of (effective Hamiltonian) predictions.
 As a result, during the comparison, we distinguish
between the experimental and predicted transitions.

\begin{figure}
\centering
{\leavevmode \epsfxsize=9.0cm \epsfbox{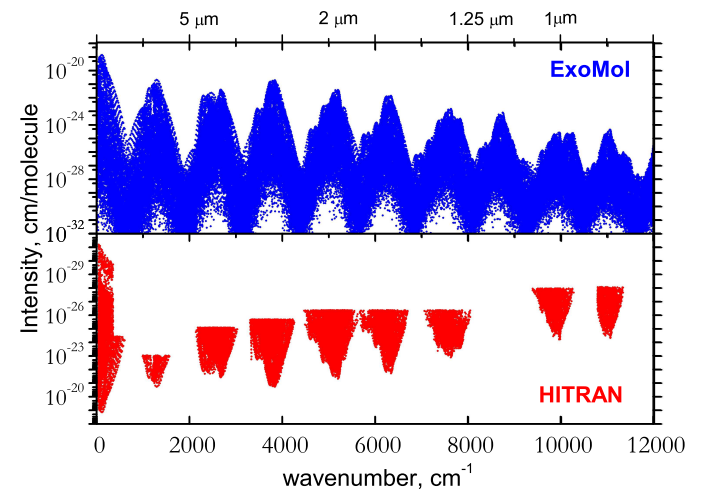}}
\caption{A $T=296$~K absorption stick spectrum of H$_2$S: Comparison with HITRAN~2012. }
\label{H2S_298K_HITRAN_12KB}
\end{figure}

\begin{figure}
\centering
{\leavevmode \epsfxsize=8.0cm \epsfbox{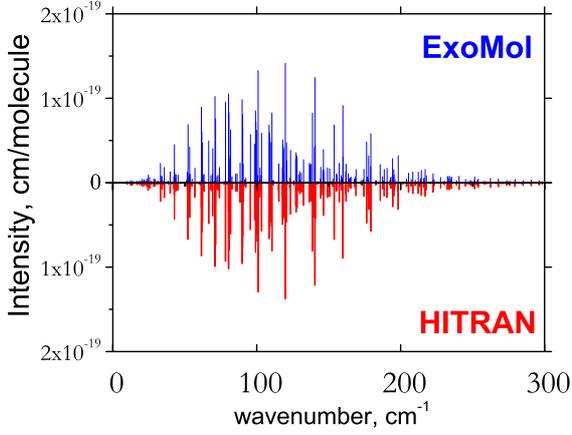}}
\caption{A calculated rotational band compared to that from the HITRAN database at $T = 296$ K.}
\label{our_calculations_compared_to_HITRAN_rotational_band}
\end{figure}

\begin{figure}
\centering




{\leavevmode \epsfxsize=4.0cm \epsfbox{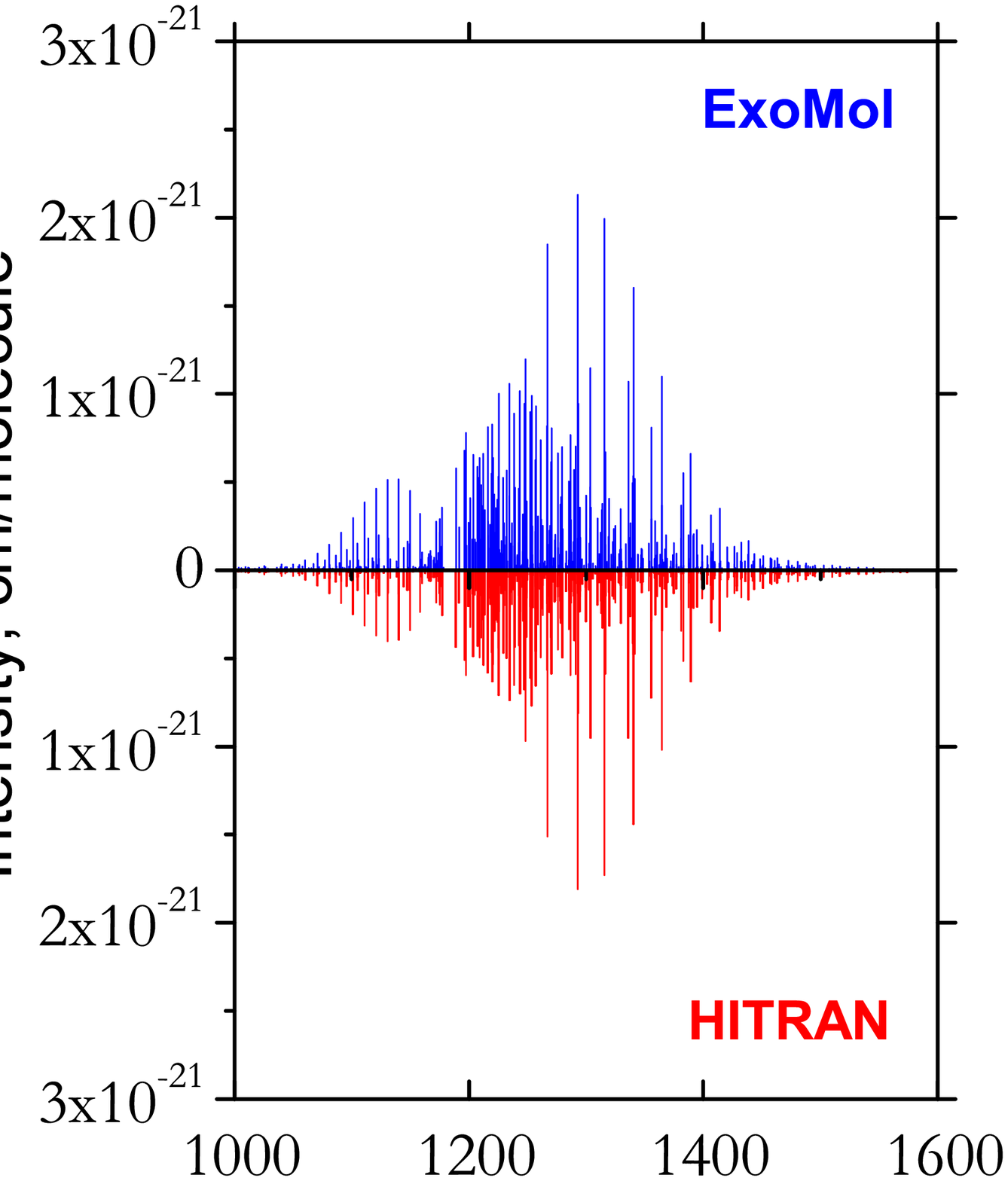}}
{\leavevmode \epsfxsize=4.0cm \epsfbox{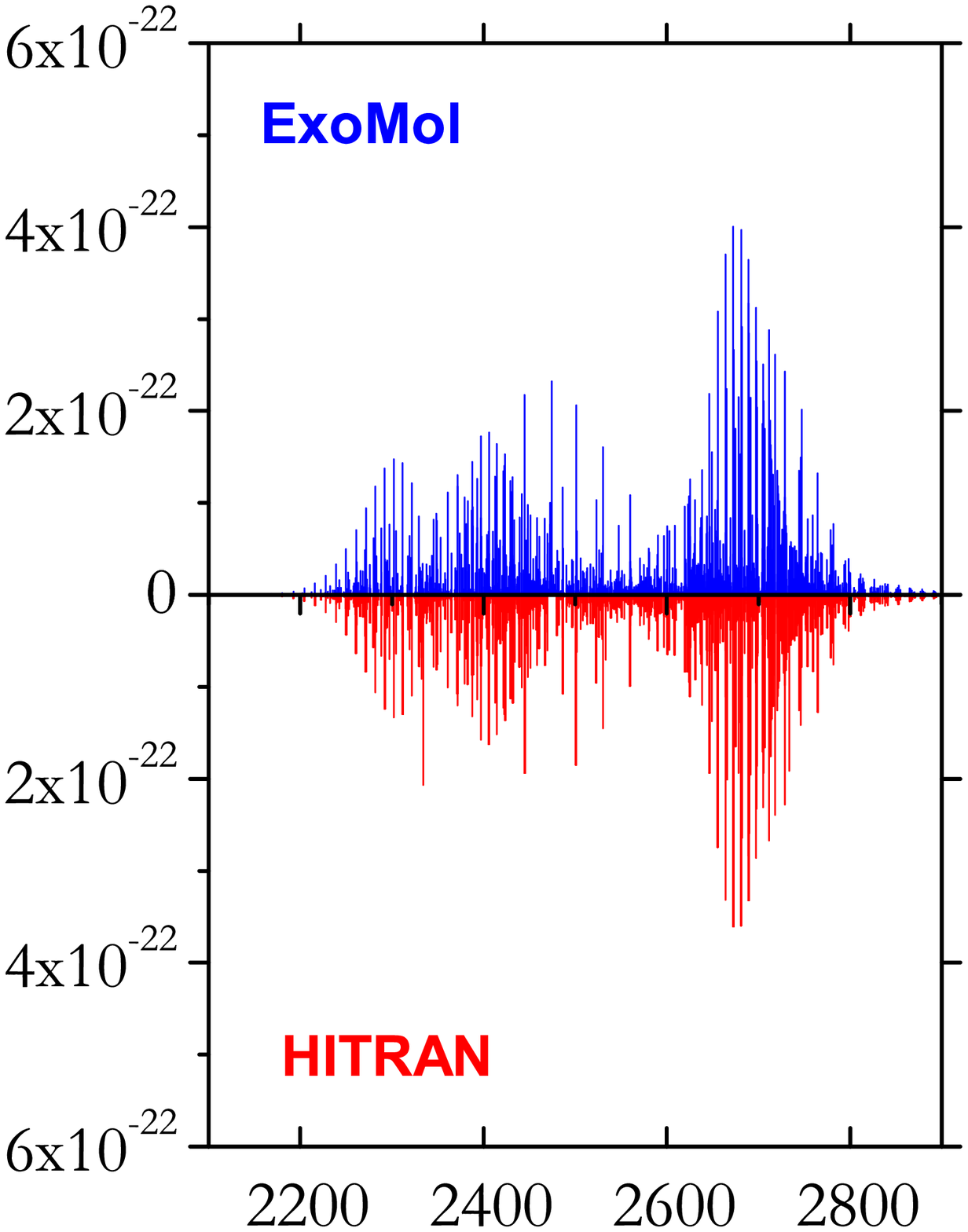}}
{\leavevmode \epsfxsize=4.0cm \epsfbox{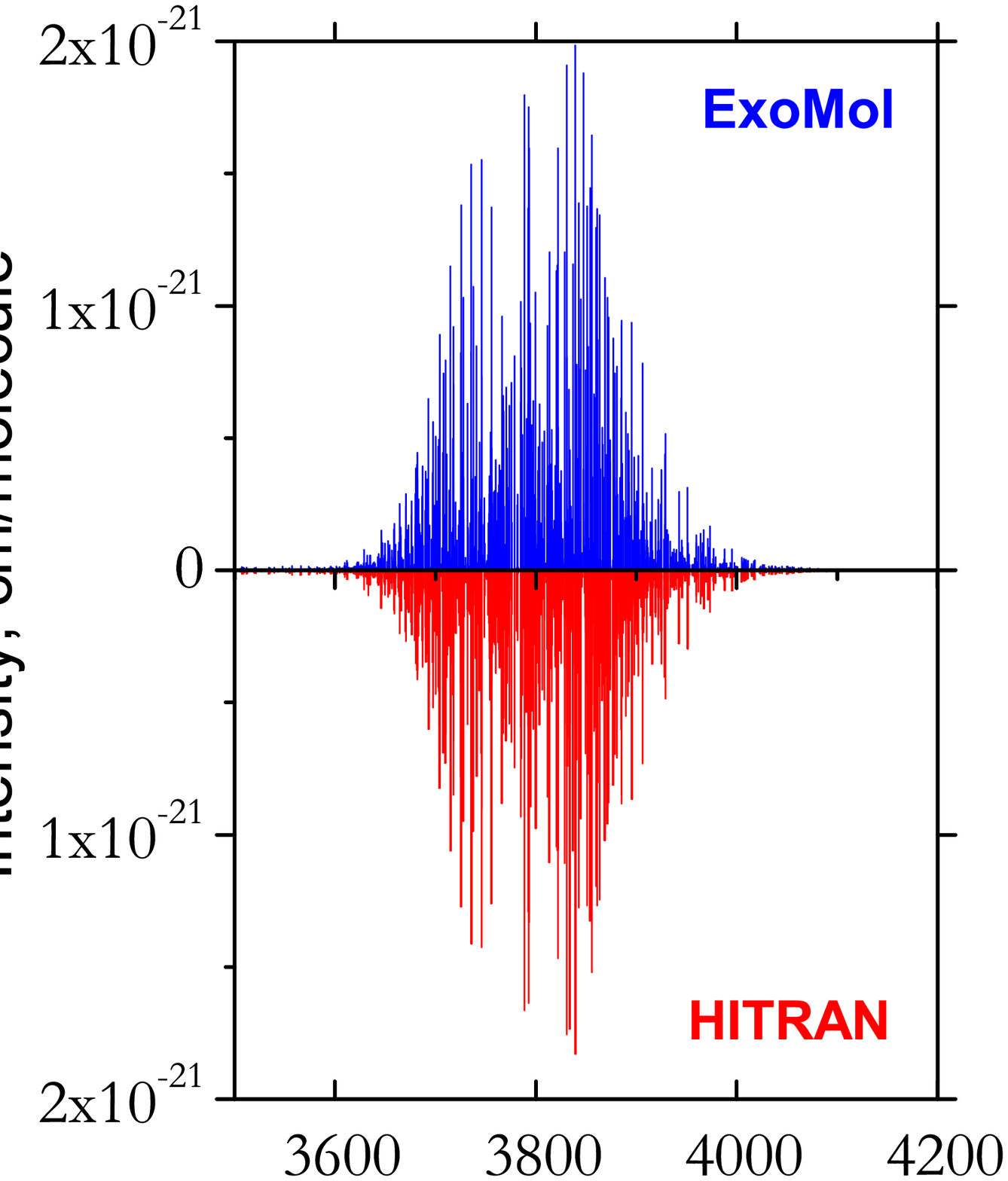}}
{\leavevmode \epsfxsize=4.0cm \epsfbox{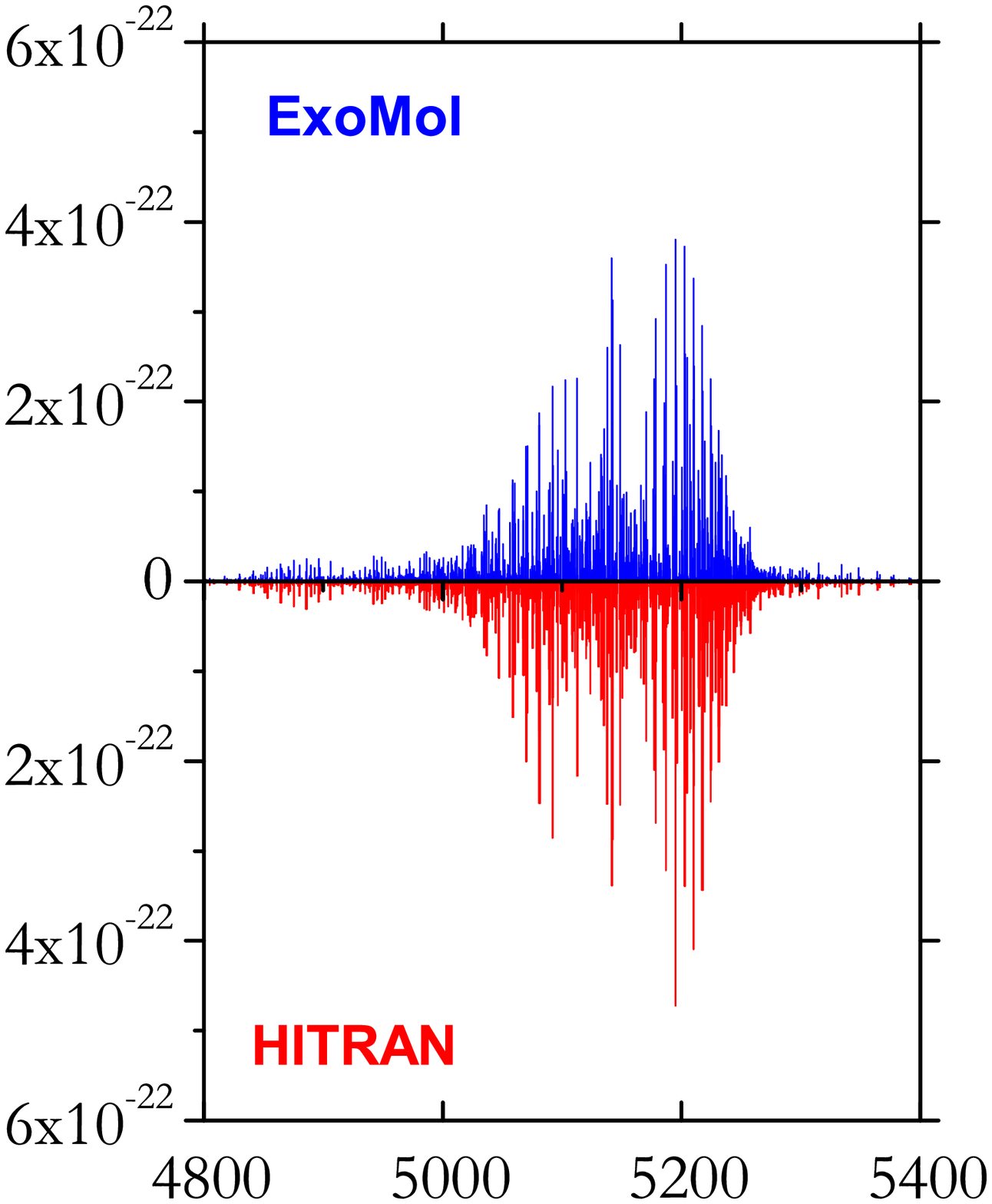}}
{\leavevmode \epsfxsize=4.0cm \epsfbox{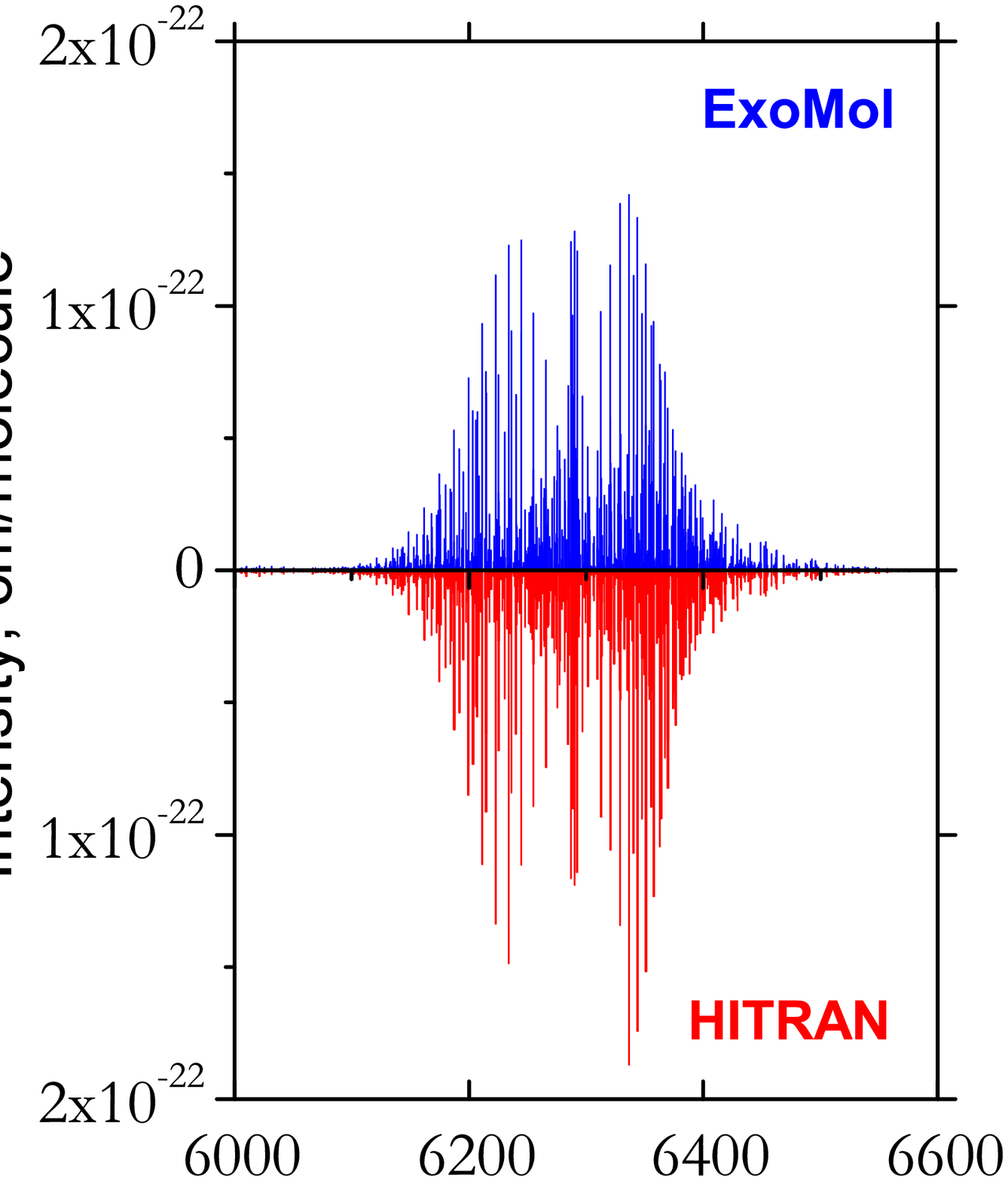}}
{\leavevmode \epsfxsize=4.0cm \epsfbox{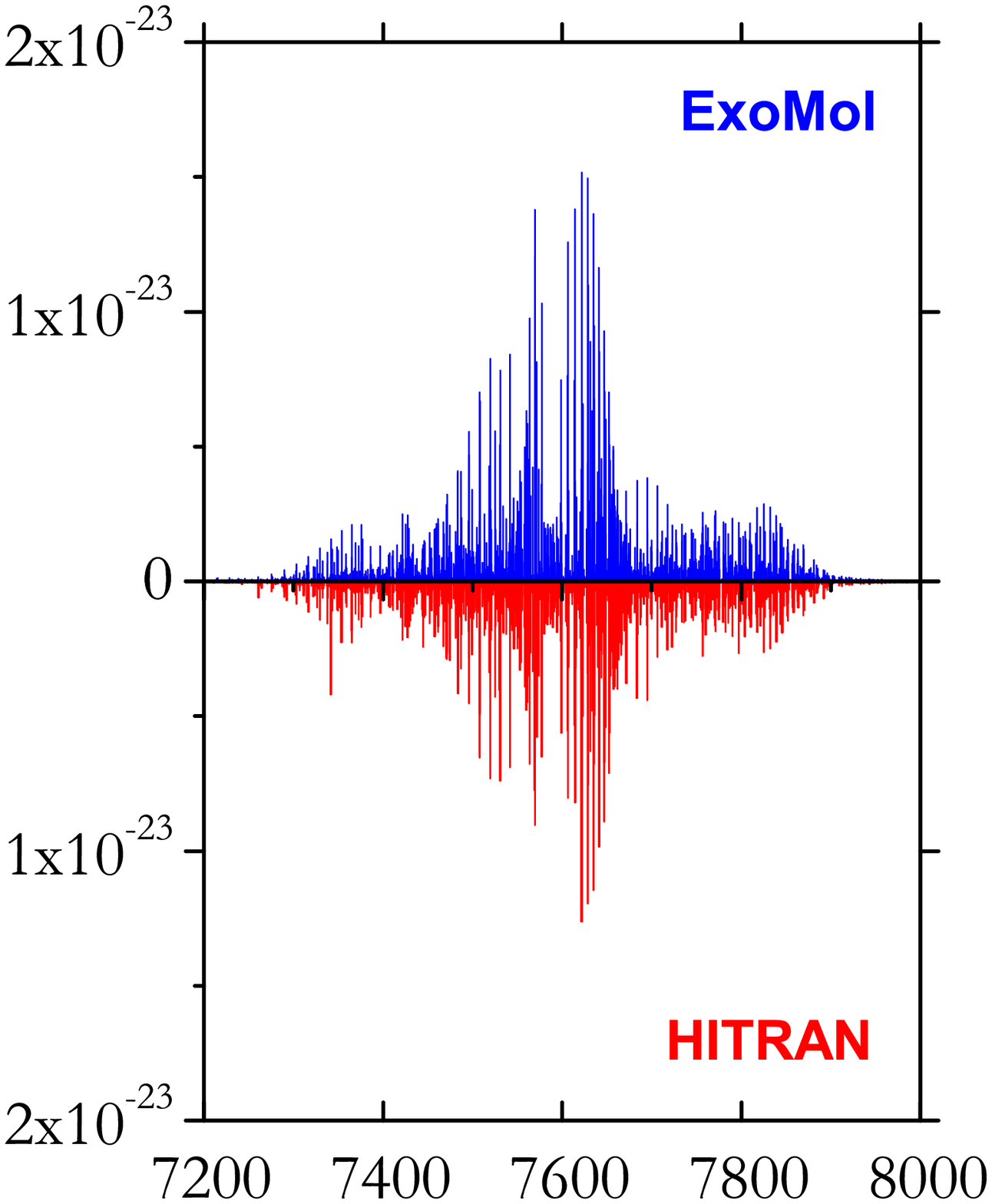}}
{\leavevmode \epsfxsize=4.0cm \epsfbox{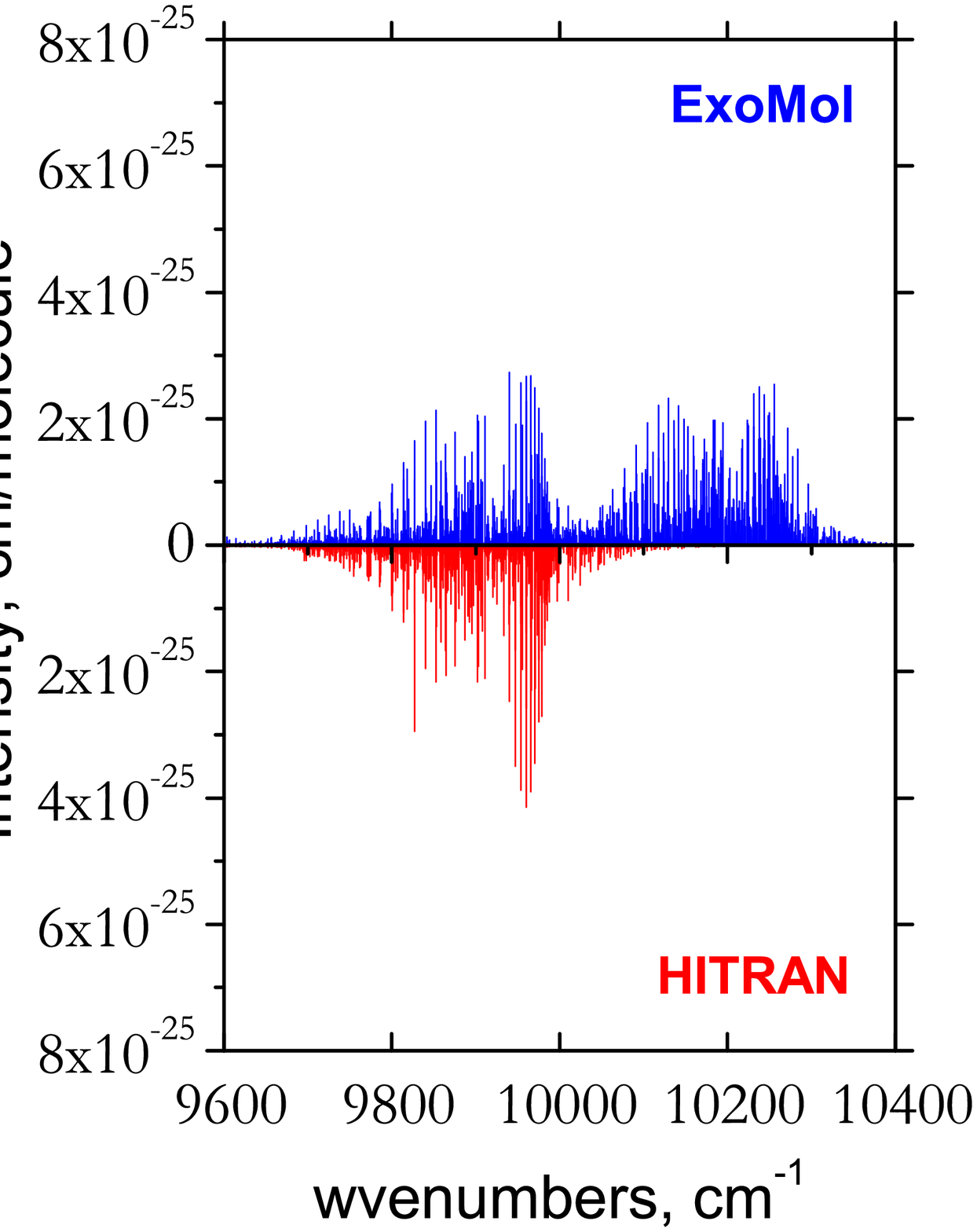}}
{\leavevmode \epsfxsize=4.0cm \epsfbox{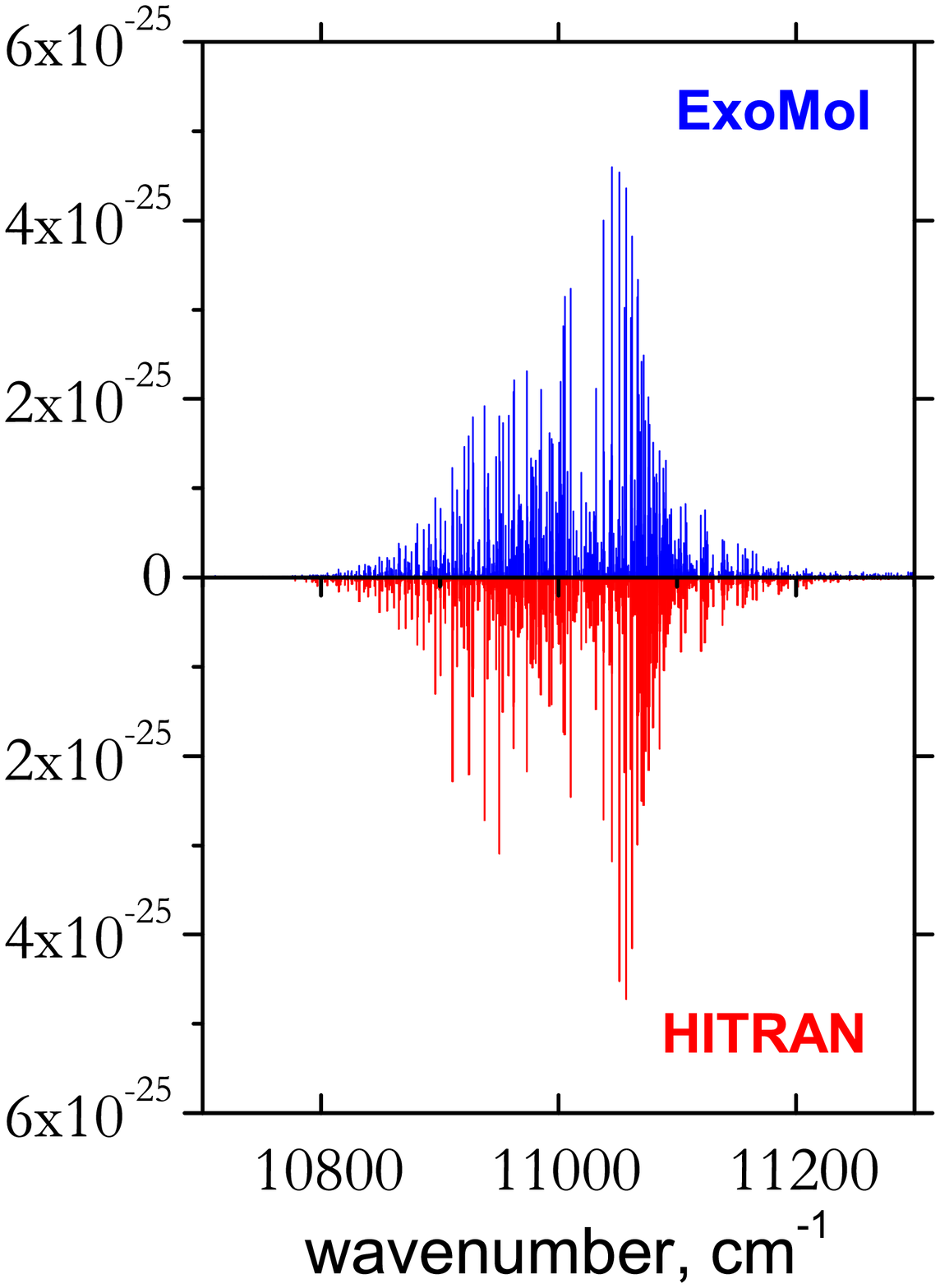}}

\caption{ Calculated spectrum at $T = 296$ K compared to that from the  HITRAN database for the polyads (top left to bottom right)
0.5, 1, 1.5, 2, 2.5,  3, 4 and 4.5.}
\label{our_calculations_compared_to_HITRAN_band}
\end{figure}

As a preliminary comparison, Fig.~\ref{H2S_298K_HITRAN_12KB} gives a general idea
about the number of transitions available in HITRAN~2012 \citep{jt557} compared to the number of
transitions calculated in this work at room temperature. Also,
Figs.~\ref{our_calculations_compared_to_HITRAN_rotational_band} and \ref{our_calculations_compared_to_HITRAN_band}
give a general idea about the accuracy of our calculated line list (ATY2) compared to the data in these databases; more detailed
comparisons are presented in the following subsections.

The data available in HITRAN 2012 for H$_2$$^{32}$S comprise 36~533 transitions in the spectral region 2- 11330 \cm\ and
covers  $J$ values up to 30 and cut-off
intensity of the order of
$10^{-32}$~cm$^{-1}$/(molecule$\times$cm$^{-2}$) at 296 K.
 Our comparisons prioritise
actual measured line positions and intensities, which we identified
from the data given in HITRAN using additional information provided by
the IS WADIS (wadis.saga.iao.ru) database.  A summary 
of the comparison between our calculations and the data in HITRAN
database for every polyad region is presented in
Tables~\ref{comprehensive_comparison_with_HITRAN-1},
\ref{comprehensive_comparison_with_HITRAN-2} and
\ref{comprehensive_comparison_with_HITRAN-3}.

\subsubsection{HITRAN database: 0--4200 \cm}
\label{HITRAN database}

	HITRAN 2012 contains around 14~700 transitions  in the 0--4250 \cm\
spectral region  with an intensity cut-off
of the order of  $10^{-32}$\ below  360 \cm\   and $10^{-26}$~cm$^{-1}$/(molecule$\times$cm$^{-2}$)
above 360 \cm.
Of all data provided by the HITRAN 2012 in this region, only around 10~700 transitions
with positions either measured or accurately determined from the upper and lower experimental
energy levels are included in our comparison. In addition, about 550 simulated transitions
of the 010--000 band between 1000 and 1570 \cm\ were also used,
as no other data are available in HITRAN 2012 for this region.
It should be noted that all intensities provided by the HITRAN below 4250 \cm\
are calculated values. Figure 5 shows the differences between ATY2 and HITRAN2012 in this region
for both transition frequencies and intensities.
A summary  of the comparison between our
calculations and the HITRAN data  for  polyad 0 to 1.5 is presented in Tables 7-8.

For the pure rotational region 0-360 \cm\ (000 - 000 and 010-010 transitions),
the standard or rms deviation  between the calculated and  measured \citep{83FlCaJo.H2S,jt558}
positions up to $J$ = 26  is 0.016 \cm\ for 1815 observed lines with maximum absolute
deviation of 0.094 \cm. All the intensities in this region are calculated using the
permanent dipole moment. Both sets of calculated intensities
provided by HITRAN and variational agree very well with an rms of 5.7\%
and average ratio of 1.04.

Positions of the 010--000 band (polyad 0.5) in HITRAN were simulated based
on the spectroscopic parameters  reported by  \citet{82LaEdGi.H2S} and then normalized
to the measured values of \citet{83Stxxxxa.H2S}. Comparison with AYT2
yields an rms and maximal deviations for line positions
of  0.049 and 0.258 \cm. Variational intensities for the 010--000
band differ significantly from a simulation adopted in HITRAN with an rms deviation of 30\%.
 Measured intensities of 103 lines of the 010--000 band reported by \citet{83Stxxxxa.H2S} ,
deviate from variational values within an rms of 23\%. The integrated variational
intensity of the 010--000 band is about 17\% larger than that estimated from  HITRAN.
This is in agreement with tests for $J \leq 5$ made previously by
\citet{jt607}.




For  the 2140-4250 \cm\ spectral region our comparison  was limited to 7505 `empirical'
transitions.  Similar rms and maximum deviations around 0.07 and 0.36 \cm,
respectively, have been obtained between 'empirical' and calculated positions
for polyads 1 and 1.5. In this comparison we used corrected set of the transition
positions above 2000 \cm\ which corresponds to the original data of \citet{98BrCrCr.H2S}.
These data are  available in the 2015 release of the GEISA-2015 database \citep{jt636}.

Above 2200 \cm\ the transition intensities adopted in HITRAN are taken from
an effective Hamiltonian  based on about 1100
accurate (within 2-5\%) measured intensities reported by  \citet{98BrCrCr.H2S}.
Comparison of the AYT2 intensities with the HITRAN data and
accurate measured values is summarized  in Table 7 and is illustrated on Fig.5.
Reasonable agreement between the measured and variational intensities within
an rms of 8\%,  is achieved, while the predicted HITRAN intensities 
deviate from the AYT2 predictions by about 16-18\% .


\begin{figure}
\centering
\includegraphics[scale=0.33]{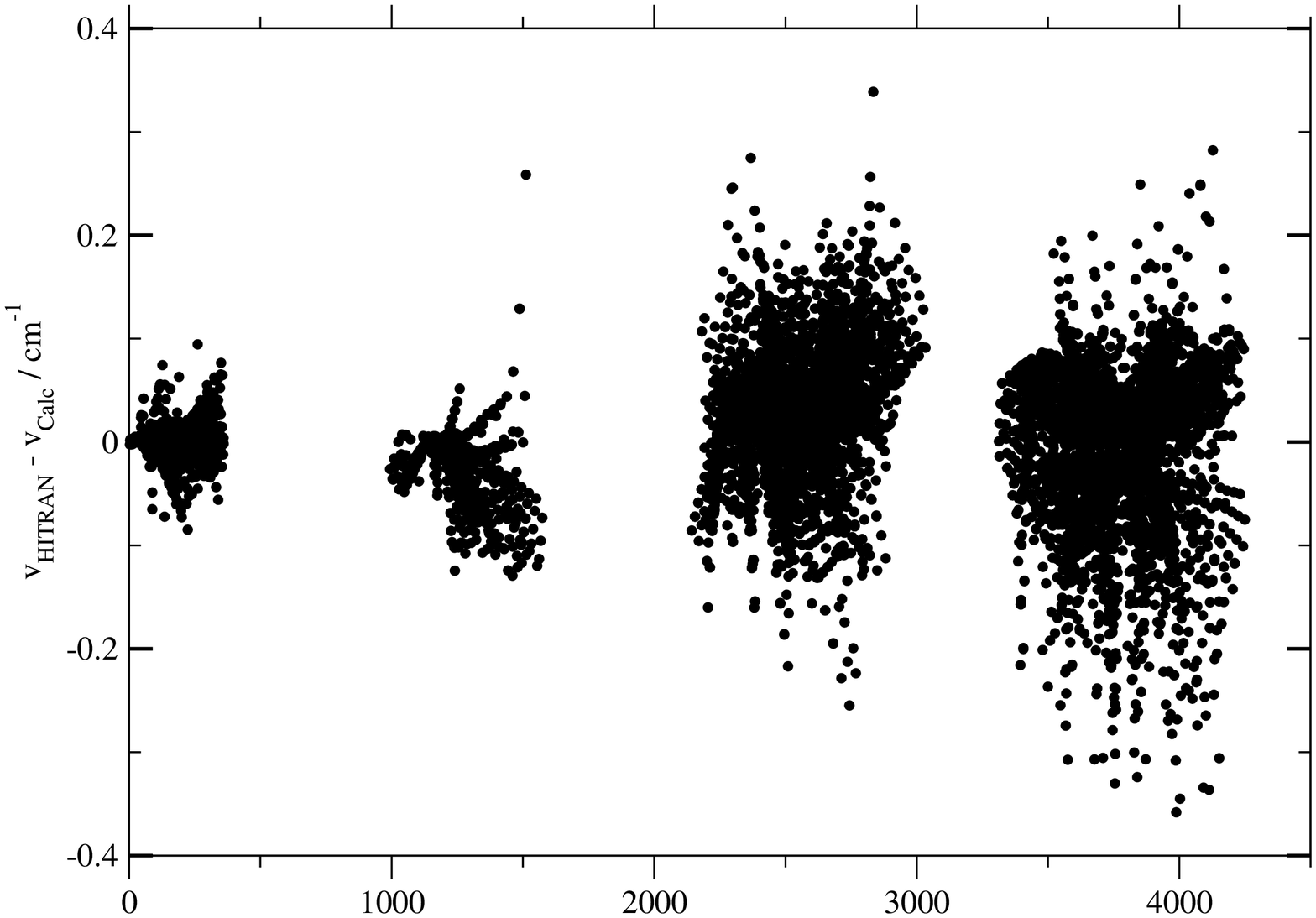}
\includegraphics[scale=0.33]{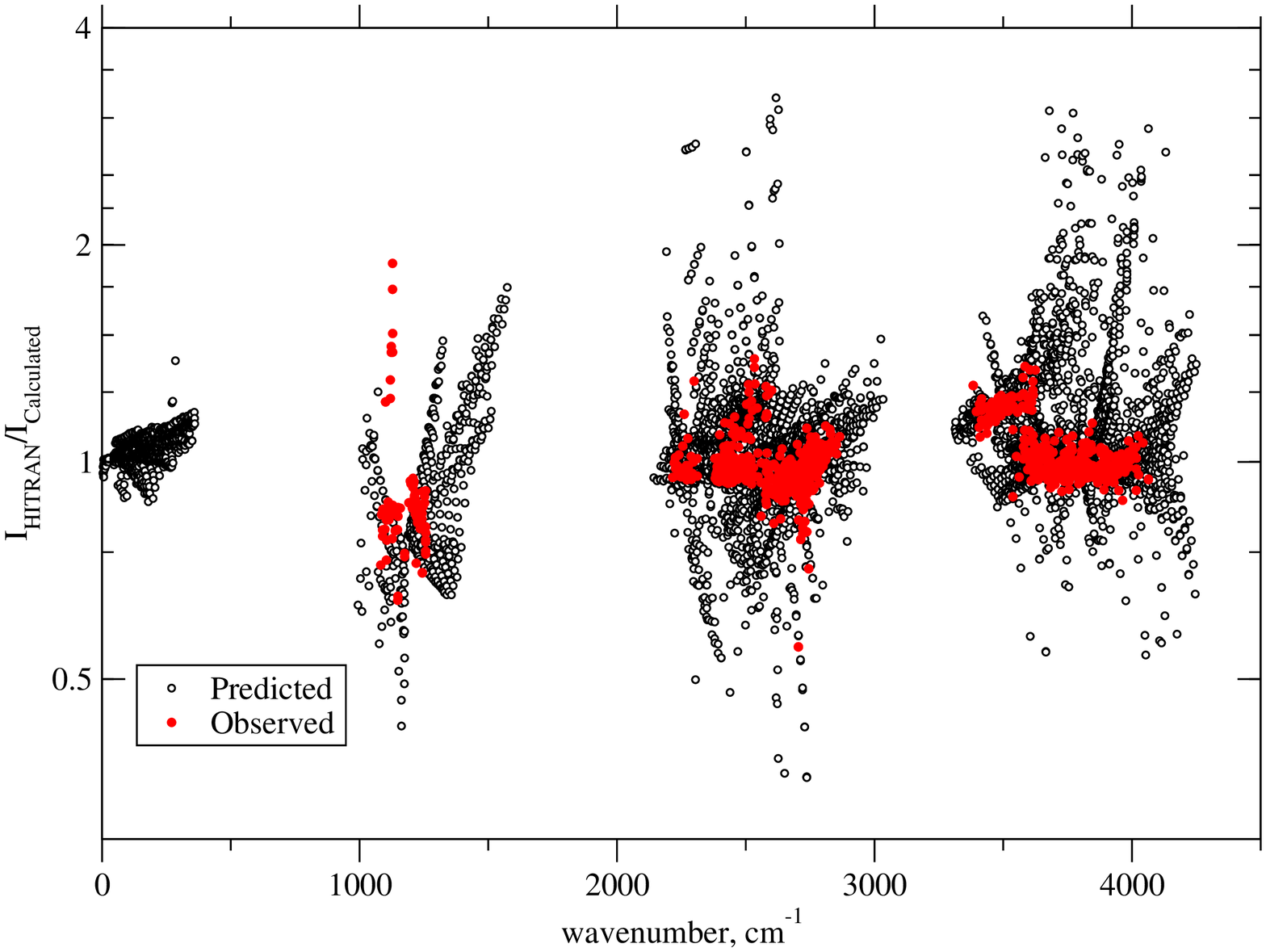}
\caption{Differences in line positions (upper) and  transition intensities (lower) compared to data available in HITRAN below 4000 \cm.
The measured and predicted lines
in this database are considered separately.}
\label{comparing-to-HITRAN}
\end{figure}

\begin{table*}
\caption{Comparison between the calculated transitions  and the data available in HITRAN for the polyad regions
0 and 0.5.
$\sigma$ is the standard deviation. Absolute intensities are in
cm$^{-1}$/(molecule$\times$cm$^{-2}$) (powers of ten in parenthesis) and errors are all relative; the frequency errors
are absolute errors. The HITRAN data are divided between directly measured and predicted from effective Hamiltonians.
}
\label{comprehensive_comparison_with_HITRAN-1}

\begin{tabular}{c l c c l c c}
\hline\hline
 Polyad     &      \multicolumn{2}{c}{Frequency}    &&       \multicolumn{3}{c}{Intensity}                                     \\
\cline{2-3}\cline{5-7}
      &                      &  Measured  &&                      & Predicted        & Measured    \\
\hline
  0   & Spectral Range       &3 -- 360 \cm   &&  Intensity Range            &1.53(-26) -- 9.91(-20) \\
      &   $J$ range          &1 -- 26     &&   Average ratio        &1.04                  \\
      &   $K_a$ range        &0 -- 17    &&   $\sigma$              &5.7 \%                \\
      &   $\sigma$           &0.016 \cm   &&   Min. ratio           &0.88                 \\
      &   Max. deviation     &0.094 \cm   &&   Max. ratio           &1.38                  \\
      & Total $\#$ lines     &1185        &&                        & 1185                     \\
      &                      &            &&                        &                      \\
 0.5  &  Spectral range      & 994 -- 1573 \cm   &&   Intensity    Range      &9.10(-24) -- 1.81(-21)       &1.52(-24) -- 8.34(-21) \\
      &   $J$ range          &1 -- 16     &&   Average ratio        &0.93 & 0.90               \\
      &   $K_a$ range        &0 -- 14    &&   $\sigma$            &30 \%  & 23 \%               \\
      &   $\sigma$           &0.016 \cm  &&   Min. ratio           &0.43 & 0.64                \\
      &   Max. deviation     &0.094 \cm  &&   Max. ratio           &1.75 & 1.88                  \\
      & Total $\#$ lines     &551        &&                        &  551 & 103                    \\
\hline\hline
\end{tabular}
\end{table*}

\begin{table*}
\caption{Comparison between the calculated transitions  and the data available in HITRAN for the polyad regions
1 and 1.5.
$\sigma$ is the standard deviation. Absolute intensities are in
cm$^{-1}$/(molecule$\times$cm$^{-2}$) (powers of ten in parenthesis) and errors are all relative; the frequency errors
are absolute errors. The HITRAN data are divided between directly measured and predicted from effective Hamiltonians.
}
\label{comprehensive_comparison_with_HITRAN-2}

\begin{tabular}{c l c c l c c}
\hline\hline
 Polyad     &      \multicolumn{2}{c}{Frequency}    &&       \multicolumn{3}{c}{Intensity}                                     \\
\cline{2-3}\cline{5-7}
      &                      &  Empirical  &&                      & Predicted        & Measured    \\
\hline
  1   & Spectral Range       &2142 -- 3034 \cm   &&  Intensity Range            &8.07(-26) -- 3.61(-22) & 5.90(-25) -- 2.73(-22)\\
      &   $J$ range          &0 -- 20     &&   Average ratio        &1.03 & 0.99                 \\
      &   $K_a$ range        &0 -- 14    &&   $\sigma$              &17.9\% & 8.2 \%                \\
      &   $\sigma$           &0.070 \cm   &&   Min. ratio           &0.36 & 0.55                 \\
      &   Max. deviation     &0.338 \cm   &&   Max. ratio           &3.18 & 1.39                  \\
      & Total $\#$ lines     &3453        &&                        & 3413 & 568                     \\
      &                      &            &&                        &                      \\
 1.5  &  Spectral range      & 3312 -- 4250 \cm   &&   Intensity    Range      &2.04(-26) -- 1.83(-21) & 4.89(-25) -- 1.77(-21)\\
      &   $J$ range          &0 -- 20     &&   Average ratio        &1.09 & 1.02                 \\
      &   $K_a$ range        &0 -- 15    &&   $\sigma$              &15.6\% & 7.5 \%              \\
      &   $\sigma$           &0.072 \cm  &&   Min. ratio            &0.54 & 0.88                 \\
      &   Max. deviation     &0.358 \cm  &&   Max. ratio            &3.06 & 1.35                   \\
      & Total $\#$ lines     &4050        &&                        & 4026 & 526             \\
\hline\hline
\end{tabular}
\end{table*}

\begin{table*}
\caption{Comparison between the calculated transitions  and the data available in HITRAN for the polyad regions
2 and 2.5.
$\sigma$ is the standard deviation. Absolute intensities are in
cm$^{-1}$/(molecule$\times$cm$^{-2}$) (powers of ten in parenthesis) and errors are all relative; the frequency errors
are absolute errors. The HITRAN data are divided between accurately and approximately measured.
}
\label{comprehensive_comparison_with_HITRAN-3}

\begin{tabular}{c l c c l c c}
\hline\hline
 Polyad     &      \multicolumn{2}{c}{Frequency}    &&       \multicolumn{3}{c}{Intensity}                                     \\
\cline{2-3}\cline{5-7}
      &                      &  Measured  &&                      & Approx. measured        & Measured    \\
\hline
2  &  Spectral range      & 4486 -- 5595 \cm   &&   Intensity    Range      &1.88(-26) -- 4.72(-22) & 4.26(-26) -- 4.09(-22)\\
      &   $J$ range          &0 -- 20     &&   Average ratio        &1.08 & 1.01                 \\
      &   $K_a$ range        &0 -- 14    &&   $\sigma$              &24.8\% & 6.9 \%              \\
      &   $\sigma$           &0.127 \cm  &&   Min. ratio            &0.50 & 0.76                 \\
      &   Max. deviation     &0.524 \cm  &&   Max. ratio            &3.00 & 1.38                   \\
      & Total $\#$ lines     &5617        &&                        & 5165 & 1423             \\
     &                      &            &&                        &                      \\
 2.5  &  Spectral range      & 5688 -- 6653 \cm   &&   Intensity    Range      &1.60(-26) -- 1.87(-22) & 2.11(-26) -- 1.87(-22)\\
      &   $J$ range          &0 -- 18     &&   Average ratio        &1.12 & 1.05                 \\
      &   $K_a$ range        &0 -- 14    &&   $\sigma$              &32.3\% & 14.0 \%              \\
      &   $\sigma$           &0.078 \cm  &&   Min. ratio            &0.40 & 0.71                 \\
      &   Max. deviation     &0.293 \cm  &&   Max. ratio            &3.00 & 1.50                   \\
      & Total $\#$ lines     &3153        &&                        & 3014 & 1019             \\
     &                      &            &&                        &                      \\
 3  &  Spectral range      & 7096 -- 7995 \cm   &&   Intensity    Range      &1.63(-26) -- 1.26(-23) \\
      &   $J$ range          &0 -- 17     &&   Average ratio        &0.99                \\
      &   $K_a$ range        &0 -- 10    &&   $\sigma$              &33.8\%               \\
      &   $\sigma$           &0.129 \cm  &&   Min. ratio            &0.50                 \\
      &   Max. deviation     &0.358 \cm  &&   Max. ratio            &3.00                    \\
      & Total $\#$ lines     &2504        &&                        & 2327             \\
     &                      &            &&                        &                      \\
 4  &  Spectral range      & 9541 -- 10001 \cm   &&   Intensity    Range      &2.98(-28) -- 6.09(-25) \\
      &   $J$ range          &0 -- 17     &&   Average ratio        &1.48                \\
      &   $K_a$ range        &0 -- 11    &&   $\sigma$              &54\%               \\
      &   $\sigma$           &0.138 \cm  &&   Min. ratio            &0.25                 \\
      &   Max. deviation     &0.670 \cm  &&   Max. ratio            &3.95                    \\
      & Total $\#$ lines     &1716        &&                        & 1568             \\
     &                      &            &&                        &                      \\
 4.5  &  Spectral range      & 10790 -- 11298 \cm   &&   Intensity    Range      &2.67(-28) -- 4.72(-25) \\
      &   $J$ range          &0 -- 19     &&   Average ratio        &1.48                \\
      &   $K_a$ range        &0 -- 11    &&   $\sigma$              &56\%               \\
      &   $\sigma$           &0.086 \cm  &&   Min. ratio            &0.27                 \\
      &   Max. deviation     &0.580 \cm  &&   Max. ratio            &2.86                    \\
      & Total $\#$ lines     &1093        &&                        & 1015             \\
\hline\hline
\end{tabular}
\end{table*}

\subsubsection{HITRAN database: 4400-8000 \cm}
\label{ midIR}

	The H$_2$S transitions in the 4400-8000 \cm\  region represent measured positions and intensities
 which are augmented with `empirical' transitions generated from
experimental energy levels and  intensities predicted
using an effective Hamiltonian.
Of 16~284 transitions included in HITRAN2012, 11~277 correspond to observed data.
It should be noted that majority of the measured intensities are of
low accuracy, often derived from a peak absorption. However, for about
2500 transitions between 4580 and 6575 \cm, accurate  intensities are provided.

Only observed data were used in comparison with variational calculation.
A summary of this comparison is given in Table 8 and Fig. 6. The observed
position are reproduced with an rms  of 0.08 \cm\ (polyad 2.5) and 0.13 \cm\
(polyads 2 and 3) with maximal deviations not exceeding 0.6 \cm.

Intensity comparisons are not so straightforward with strongly distorted
ratios between observed and AYT2 intensitions for individual transitions. Sometimes this disagreement is due
to missing additional lines, or due to incorrect assignment, sometimes the
experimental line is a superposition with another H$_{2}$S isotopologue, removed from the list.
There is also the possibility of increased errors in the AYT2 intensities for individual
vibrational bands, due to problems with the DMS, or for individual transitions, due to
perturbations \citep{jt522,jt625}. The distribution
of intensity ratios was examined for every polyad, and the suitable upper and lower
limits chosen which included 91 - 95\%\ of the lines; these were used to determine
the rms deviations presented in Table 8.

The best intensity agreement, with an rms of 6.9\%, was obtained for accurate 1423
experimental intensities for polyad 2, while the rms for all observed lines
for this polyad was about 25\%. At the same time, intensities of transitions belonging
to polyad 2.5 deviated from their AYT2 analogs,
see Table 8 and Fig.6, both for approximate and accurate measured data: of 1084 accurate
intensities only 1019 could be reproduced with an rms of  14\%.
We note that our calculation
appear to overestimate the strength of the $5\nu_2$ band, which lies
at about 5800~\cm, by approximately a factor of 2. This is almost
certainly caused by residual problems with DMS of \citet{jt607}.

\subsubsection{HITRAN database: 9500-11300 \cm}
\label{IAO LMS Spectra database}


The HITRAN2012 data for H$_{2}$S in this region comprises the  measurements of
\citet{01NaCaxxb.H2S} and \citet{01NaCaxxa.H2S}
augmented by the calculated lines with `empirical' positions
and intensities predicted on the basis of and effective Hamiltonian model.
Of 5605 transitions, only 2835 have measured positions and intensities.
Experimental intensities  were
estimated to be accurate to 25-30\% on average and up to 100\% for the weakest lines in polyad 4 region,
and as 15\% for strong and medium intensity lines and up to 50\% for weakest lines for polyad 4.5.

Again only measured transitions were used in the comparison.
Table~\ref{comprehensive_comparison_with_HITRAN-3} summarizes the results
of the analysis which are shown graphically in
Fig. \ref{comparison-with-spectra-databaseintens}.
Agreement between the observed and AYT2 line positions was found to be  quite satisfactory
with an rms of 0.086 and  0.138 \cm\ for polyad 4.5 and 4, respectively,
while the maximal deviation does not exceed 0.67 \cm\ for both polyads.
However, the intensities agree less well than for the lower
polyads with an rms of 54-56\% when about 90\% of lines were compared.
Much larger deviations in intensity ratios (up to 2 orders of magnitude) were also encountered.
They may be caused by the low accuracy of the experimental intensities,
incomplete or incorrect assignments, or by distortion of the calculated values.
It seems that calculated intensities of transitions involving the upper states in the local mode limit are strongly
sensitive to the details of the PES, and can change significantly for its different versions.



In general, our calculations agree well with the measured transition
intensities where we find the difference of up to
about a factor of 3 while much larger differences sometimes up to
two-three orders of magnitudes appear for results
derived from effective Hamiltonians (not shown). This analysis suggests that
there are problems with the effective Hamiltonian extrapolation; this problem
will be analysed elsewhere.

Our calculated spectrum at room temperature contains around 620~000
transitions up to $J$~=~40 below 12~000~\cm\ with the intensity
cut-off $10^{-31}$~cm$^{-1}$/(molecule$\times$cm$^{-2}$) comparing to
36~600 transitions in HITRAN 2012. This spectrum has a standard
deviation for the transition positions of about 0.072 \cm\ for 94\% of
the 20~513 most accurate measured lines below 8000 \cm\ (see Tables
7-9). The standard deviation of the ratios of the transition
intensities in HITRAN 2012 to the calculated intensities is 10\% with
an average ratio of 1.02 for 76\% of lines considered, provided that
the less accurate 010--000 transition intensities were excluded from
consideration.


\begin{figure}
\centering
\includegraphics[scale=0.33]{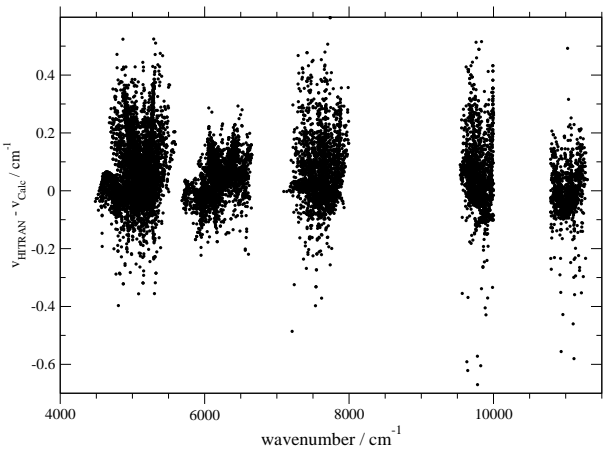}
\caption{Differences in line positions comparing to actual observed data included in HTRAN2012.}
\label{comparison-with-spectra-databasefrreq}
\end{figure}

\begin{figure}
\centering
\includegraphics[scale=0.33]{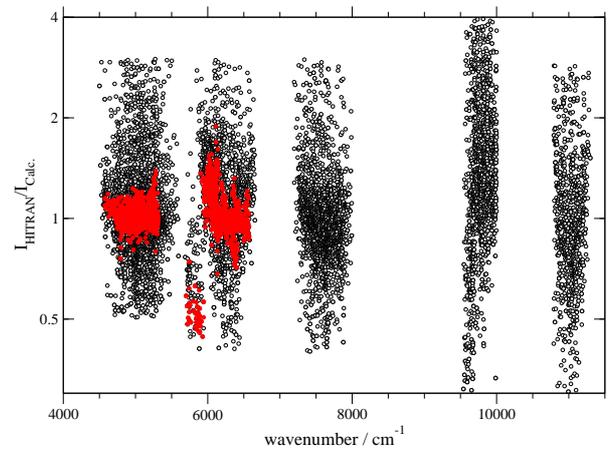}
\caption{Errors in transition intensities comparing to  actual observed data included in HTRAN2012;
intensities for which the measurements are accurate are highlighted.}
\label{comparison-with-spectra-databaseintens}
\end{figure}

\section{Partition function and lifetimes}
\label{Partition function}

\begin{figure}
\centering
\includegraphics[scale=0.33]{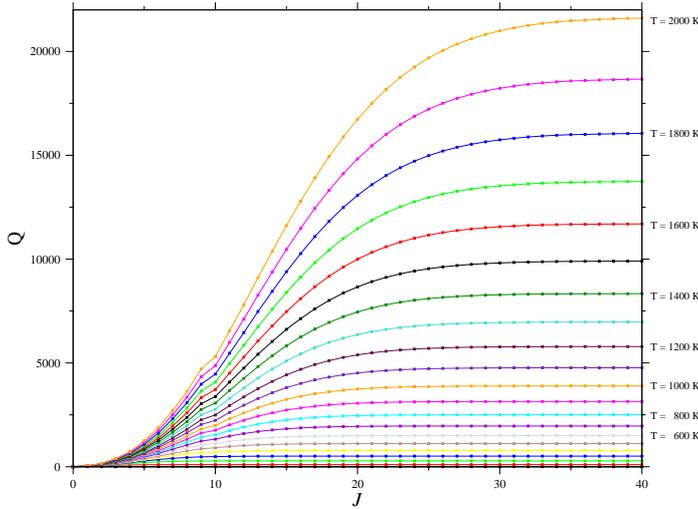}
\caption{Partition functions convergence curves.}
\label{Partiion_function_plot}
\end{figure}

The partition function value ($Q$) of a molecule  at  temperature ($T$) is given by
\begin{equation}
\label{Partition-function-equation}
Q(T) = \sum_{J}\sum_{i}g_{i}(2J + 1) \exp\left(-\frac{c_{2}E_{i}^{(J)}}{T}\right),
\end{equation}
where $g$ is the nuclear statistical weight which is 1 or 3 for
para or ortho states of H$_{2}$$^{32}$S in the convention
adopted by HITRAN and ExoMol; $E_{i}^{(J)}$
is the ro-vibrational energy for a given $J$ value, and
$c_{2}$ is the second radiation constant.  The summation in this
equation should be performed over all ro-vibrational energy levels with
or at least until the value of $Q$ converges.
Figure~\ref{Partiion_function_plot} shows how $Q$ converges with $J$ at
different temperatures up to 2000~K.
Table~\ref{Partition_function_table} presents H$_2$$^{32}$S $Q$ values at different
temperatures in comparison with the values from the HITRAN, the
JPL \citep{98PiPoCo}, and the CDMS databases.

\begin{figure}
\centering
{\leavevmode \epsfxsize=8.0cm \epsfbox{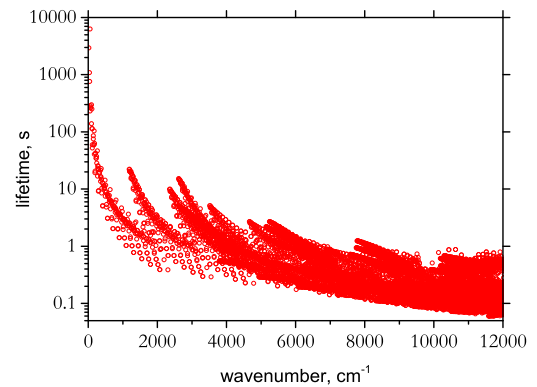}}
\caption{Lifetimes of H$_2$S obtained using the line list (ATY2).}
\label{lifetimes}
\end{figure}

The lifetimes of H$_2$S were computed using the Einstein~A coefficients from the ATY2 linelist following
the methodology laid out by \citet{jt631}. Results are summarisd in
Fig.~\ref{lifetimes}; we note that the lifetimes are much more regular as function of changes
in quantum numbers than those computed by \citep{jt631} for water.

\begin{table}
\caption{H$_2$$^{32}$S partition function values at different temperatures from different sources.}
\label{Partition_function_table}
\scalebox{0.9}{
\begin{tabular}{r r r r r r}
\hline\hline
\multicolumn{1}{c}{$T$}&\multicolumn{1}{c}{This work}&\multicolumn{1}{c}{\rm HITRAN$^{a}$}   &\multicolumn{1}{c}{JPL$^{b}$}\\
\hline
         2.725 &             1.0077 &                    &                    \\
             5 &             1.2458 &                    &                    \\
         9.375 &             2.9106 &                    &             2.9106 \\
         18.75 &             8.6997 &                    &             8.6996 \\
          37.5 &            23.8654 &                    &            23.8654 \\
            75 &            65.5266 &            65.4854 &            65.5265 \\
           100 &           100.1557 &           99.97209 &                    \\
           150 &           182.7646 &           182.2696 &           182.7622 \\
           200 &           280.6278 &           279.6049 &                    \\
           225 &           334.6933 &           333.3219 &           334.5222 \\
           296 &           505.7921 &      503.07$^{c}$  &                    \\
           300 &           516.1942 &           513.3829 &           514.4470 \\
           400 &           803.3417 &           797.6368 &                    \\
           500 &          1146.2892 &            1136.46 &                    \\
           600 &          1552.6512 &           1537.079 &                    \\
           700 &          2032.2186 &           2009.062 &                    \\
           800 &          2596.5908 &           2563.217 &                    \\
           900 &          3258.9381 &           3211.929 &                    \\
          1000 &          4033.8404 &           3968.802 &                    \\
          1100 &          4937.2034 &           4848.387 &                    \\
          1200 &          5986.2329 &           5866.361 &                    \\
          1300 &          7199.4423 &           7039.470 &                    \\
          1400 &          8596.6730 &           8385.471 &                    \\
          1500 &         10199.1148 &           9923.019 &                    \\
          1600 &         12029.3175 &           11672.66 &                    \\
          1700 &         14111.1872 &           13655.08 &                    \\
          1800 &         16469.9578 &           15893.10 &                    \\
          1900 &         19132.1349 &           18409.06 &                    \\
          2000 &         22125.4048 &           21229.28 &                    \\
\hline\hline
\end{tabular}
}

\noindent
$^{a}$ As calculated using the HITRAN's FORTRAN programs for partition functions sums.
 \\
$^{b}$ As published on the JPL's website; values are very similar to those given by CDMS.\\
$^{c}$ From \citet{06SiJaRo.method}.
\end{table}

\section{Hot spectra}
\label{Hot linelist}

\begin{figure}
\centering
{\leavevmode \epsfxsize=9.5cm \epsfbox{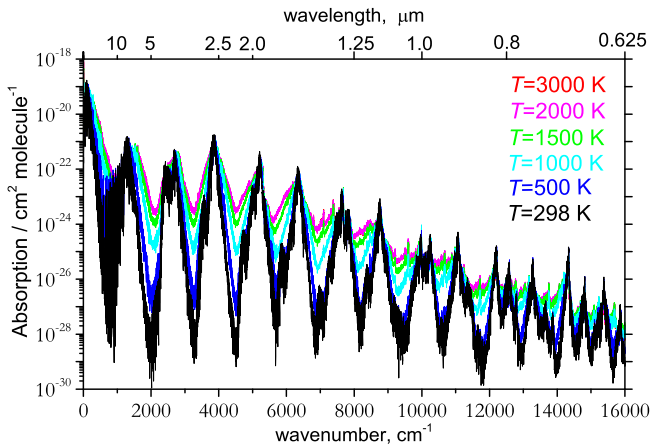}}
\caption{Temperature-dependent spectra generated using the  ATY2 line list.}
\label{our_spectrum_at_different_tempretures}
\end{figure}

Fig.~\ref{our_spectrum_at_different_tempretures} shows the calculated
spectra at different temperatures.  This figure shows how the weak
transitions at low temperatures become stronger at higher
temperatures.  The spectrum at 2000~K contains 46 million 
transitions up to 12~000~\cm, with $J \le$ 40 with intensities below 
$10^{-31}$~cm$^{-1}$/(molecule$\times$cm$^{-2}$).  Transition
positions do not change with temperature unlike transition
intensities which are temperature dependent due to the
$e^{-E_\mathrm{low}/kT}$ Boltzmann factor.  The ATY2 line list can also be
used to simulate room temperature spectra up to 20000 \cm; however,
our calculations are less accurate at visible wavelengths,

No experimental line intensities of H$_2$S can be found to compare our calculations with, neither
experimental nor theoretical with temperatures higher than $296$~K. This made our line list as the first available source of
data for H$_{2}$$^{32}$S spectrum at temperatures higher than the room
temperature, thus, opening the door for probability of identifying
H$_{2}$$^{32}$S transitions in exoplanet and brown dwarfs atmospheres. This line list can be used in the high
temperature laboratory spectra analysis.

Inspection of Fig.~\ref{our_spectrum_at_different_tempretures} shows
that, unusually, the maximum infrared absorption lies at 2.5~$\mu$m
rather than at longer wavelengths. This is a manifestation of the
unusual dipole moment of H$_2$S which results in the fundamental bands
being weaker than the overtones at about 2.5~$\mu$m.

\section{Conclusion}

A new hot line list for H$_2$S, called ATY2, has been computed
containing 114 million transitions. The line list is divided into an
energy file and a transitions file. This is done using the new
ExoMol format \citep{jt631}.  The full line list can be
downloaded from the CDS, via
\url{ftp://cdsarc.u-strasbg.fr/pub/cats/J/MNRAS/xxx/yy}, or
\url{http://cdsarc.u-strasbg.fr/viz-bin/qcat?J/MNRAS//xxx/yy}, as well
as the ExoMol website, \url{www.exomol.com}.  The line lists and
partition function together with auxiliary data including the
potential parameters and dipole moment functions, as well as the
absorption spectrum given in cross section format \citep{jt542}, can
all be obtained also from \url{www.exomol.com} as part of the extended
ExoMol database \citep{jt631}.

\section*{Acknowledgements}

This work was supported  by the ERC
under the Advanced Investigator Project 267219 and the University of Jordan.
JT and SNY thank the support of the COST action MOLIM (CM1405).
We thank the Royal Society for supporting a visit by OVN to London.

\bibliographystyle{mnras}

\label{lastpage}

\end{document}